\documentclass[twocolumn,prd,superscriptaddress,preprintnumbers,nofootinbib]{revtex4-2}

\usepackage{graphicx}
\usepackage{amsmath,bm,amssymb,amsfonts,dsfont}
\usepackage[usenames,dvipsnames]{xcolor}
\usepackage[normalem]{ulem}
\usepackage{url}
\usepackage{array}
\usepackage{booktabs}
\usepackage{multirow}
\usepackage{float}
\usepackage[colorlinks  = true,
            linkcolor   = NavyBlue,
            urlcolor    = NavyBlue,
            citecolor   = NavyBlue,
            anchorcolor = NavyBlue]{hyperref}
\usepackage{cprotect}

\usepackage{appendix}

\usepackage[switch]{lineno}

\pdfoutput=1 % if your are submitting a pdflatex (i.e. if you have
\newcommand{\lsim}{\mathrel{\mathop{\kern 0pt \rlap
  {\raise.2ex\hbox{$<$}}}
  \lower.9ex\hbox{\kern-.190em $\sim$}}}
\newcommand{\gsim}{\mathrel{\mathop{\kern 0pt \rlap
  {\raise.2ex\hbox{$>$}}}
  \lower.9ex\hbox{\kern-.190em $\sim$}}}

\begin{document}

%\preprint{CP3-XXX}
%\preprint{CTPU-PTC-24-31, CERN-TH-2024-164}

\title{Antinuclei from B-hadrons}
\title{Cosmic Antinuclei from the Decays of Beauty Baryons}
\title{Consistent Evaluation of Antinuclei Rates from the Decays of Beauty Baryons}
\title{Reliable Determination of Antinuclei Rates from the Decays of Beauty Baryons}
\title{Consistent Theoretical Analysis of Antinuclei Rates from the Decays of Beauty Baryons}
\title{Systematic Assessment of Antinuclei Production from the Decays of Beauty Baryons}
\title{Antinuclei Production from the Weakly-Decaying Beauty Hadrons}
\title{A Robust Determination of Antinuclei Production from Dark Matter via Weakly Decaying Beauty Hadrons}

\author{Mattia Di Mauro}
\email{dimauro.mattia@gmail.com}
\affiliation{Istituto Nazionale di Fisica Nucleare, Sezione di Torino, Via P. Giuria 1, 10125 Torino, Italy}

\author{Adil~Jueid}
\email{adiljueid@ibs.re.kr}
\affiliation{Particle Theory and Cosmology Group, Center for Theoretical Physics of the Universe, Institute for Basic Science (IBS), Daejeon, 34126, Republic of Korea}

\author{Jordan~Koechler}
\email{jordan.koechler@gmail.com}
\affiliation{Istituto Nazionale di Fisica Nucleare, Sezione di Torino, Via P. Giuria 1, 10125 Torino, Italy}

\author{Roberto Ruiz de Austri}
\email{rruiz@ific.uv.es}
\affiliation{Instituto de F\'{\i}sica Corpuscular, CSIC-Universitat de Val\`encia, E-46980 Paterna, Valencia, Spain}

\begin{abstract}

Recently, the Alpha Magnetic Spectrometer (\textsf{AMS-02}) Collaboration presented tentative evidence for the detection of cosmic antihelion-3 (${}^3\overline{\rm He}$) events, alongside a comparable number of antideuterons ($\overline{\rm D}$). If confirmed, these observations could revolutionize our understanding of cosmic-ray production and propagation and/or serve as compelling indirect evidence for dark matter. Given that the detection of cosmic $\overline{\rm D}$ is already at the limit of \textsf{AMS-02} sensitivity, explaining the observation of ${}^3\overline{\rm He}$ even within the standard coalescence framework poses a significant challenge.
It has recently been shown that a previously overlooked mechanism within the Standard Model of particle physics—namely, the production of antihelion via the displaced-vertex decay of $\bar{\Lambda}_b^0$ baryons—could substantially enhance the ${}^3\overline{\rm He}$ flux arising from dark matter–induced processes.
In light of these challenges, we present a tuning of \textsf{Pythia} that is consistent with \textsf{LEP} data on the fragmentation function of $b$ quarks into $b$-hadrons—a critical factor for determining the $\bar{\Lambda}_b^0$ multiplicity—and with \textsf{ALICE} and \textsf{ALEPH} data for the $\overline{\rm D}$ and ${}^3\overline{\rm He}$ spectra, which we employ to determine our coalescence model. Our refined \textsf{Pythia} tuning, in conjunction with our coalescence model, results in a predicted branching ratio for the production of ${}^3\overline{\rm He}$ from $\bar{\Lambda}_b^0$ decays that is consistent with the recent upper limit measured by \textsf{LHCb}. Furthermore, our prediction indicates that the contribution of $\overline{\rm D}$ and ${}^3\overline{\rm He}$ from beauty-hadron decays is negligible relative to the direct production from hadronization. 

\end{abstract}

%\keywords{Dark Matter, Indirect Detection experiments, Monte Carlo event generators.}
%{\let\thefootnote\relax
%\footnotetext{$^\dagger$\,Contact authors} }

\maketitle

\flushbottom

{\it Introduction--} 
The particle nature of dark matter (DM) remains unknown despite decades of extensive theoretical and experimental efforts. Well-motivated DM models have spurred a comprehensive search program that includes indirect detection, direct detection, and collider experiments \cite{Cirelli:2024ssz}. Indirect detection seeks to identify DM signals through the flux of cosmic messengers—such as positrons, antiprotons, $\gamma$ rays, and neutrinos \cite{Gaskins:2016cha,Fermi-LAT:2016afa}—although these fluxes are frequently dominated by astrophysical sources, complicating the identification of any DM contribution (see, e.g., Refs.~\cite{DiMauro:2015jxa,DiMauro:2015tfa,Genolini:2021doh,DiMauro:2021qcf,McDaniel:2023bju,Balan:2023lwg}).

Cosmic antinuclei from DM annihilation or decay offer a promising alternative. In particular, antideuterons ($\overline{\rm D}$) \cite{Donato:1999gy} and, to a lesser extent, antihelions (${}^3\overline{\rm He}$) \cite{Cirelli:2014qia,Carlson:2014ssa} are attractive search channels because their astrophysical backgrounds are significantly suppressed for kinetic energies ($K$) below 1 GeV/nucleon. In scenarios where Weakly Interacting Massive Particles (WIMPs) annihilate in the Galactic halo, the predicted $\overline{\rm D}$ flux at $K = 0.1$–1 GeV/nucleon exceeds that from secondary production by at least one order of magnitude (see, e.g., Refs.~\cite{Donato:1999gy,Ibarra:2012cc,Fornengo:2013osa,Herms:2016vop,Korsmeier:2017xzj}). Thus, the observation of even a few low-energy cosmic $\overline{\rm D}$ events could serve as a compelling DM signature \cite{vonDoetinchem:2015yva}.

So far, no firm detection of cosmic $\overline{\rm D}$ has been reported. The strongest limit comes from the \textsf{BESS} experiment, which sets an upper limit of $6.7 \times 10^{-5}$ (m$^2$ s sr GeV/n)$^{-1}$ for $K = 0.163$–1.100 GeV/n \cite{PhysRevLett.132.131001}. Nevertheless, experiments such as the Alpha Magnetic Spectrometer (\textsf{AMS-02}) aboard the International Space Station \cite{2008ICRC....4..765C} and the upcoming General AntiParticle Spectrometer (\textsf{GAPS}) mission \cite{Aramaki:2015laa} are expected to improve antinuclei detection sensitivity significantly, reaching levels as low as about $2\times 10^{-6}$ (m$^2$ s sr GeV/n)$^{-1}$ for $K < 1$ GeV/n \cite{vonDoetinchem:2015yva}.

\textsf{AMS-02} has tentatively detected about a dozen ${}^3\overline{\rm He}$ and ${}^4\overline{\rm He}$ events, as well as a few candidates with masses consistent with $\overline{\rm D}$ \cite{Tingcern2016,Miapp2022DbarHebar,Miapp2022Dbar}. The observation of roughly similar fluxes for $\overline{\rm D}$, ${}^3\overline{\rm He}$ and ${}^4\overline{\rm He}$ is unexpected since kinematic constraints imply that the formation probability drops drastically for each additional antinucleon added in the nucleus. For instance, using our coalescence model implemented in \textsf{Pythia 8}, we obtain for a 50 GeV DM candidate annihilating into $b\bar{b}$ the following multiplicity ratios:
\begin{equation}
\overline{p} : \overline{\text{D}} : {}^3\overline{\text{He}} \sim 1 : 1.4\times10^{-4} : 3.4\times10^{-8}.
\end{equation}

The authors of Ref.~\cite{Winkler:2020ltd} have proposed that a significant fraction of the ${}^3\overline{\rm He}$ flux from DM annihilation could result from the decays of $\bar{\Lambda}^0_b$ baryons, which are predominantly produced in channels involving $b\bar{b}$ quarks. Their decays efficiently produce multi-antinucleon states with small relative momenta, favoring the formation of antideuterons and, in particular, antihelions. In addition to $\bar{\Lambda}^0_b$, other weakly-decaying $b$-baryons—such as $\Sigma^0_b$, $\Sigma^{\pm}_b$, $\Xi_b^0$, $\Xi_b^-$, and $\Omega^-_b$—are produced in $e^{\pm}$ and $pp$ collisions with multiplicities that are approximately a factor of 10 smaller with respect to the $\bar{\Lambda}^0_b$.

In Ref.~\cite{Winkler:2020ltd}, a specific tune of the Monte Carlo (MC) particle generator \textsf{Pythia} (hereafter the {\bf WL21 tune}) was used to match the fragmentation function $f\left(b \rightarrow \Lambda^0_b\right)$ measured at \textsf{LEP}. This parameter is paramount for accurately modeling the rate of antinuclei production from the decays of $b$–baryons. Their approach increases the predicted yield of ${}^3\overline{\text{He}}$ from weakly decaying $b$-baryons by about a factor of 10 compared to the default \textsf{Pythia} settings, suggesting an Earth-bound ${}^3\overline{\text{He}}$ flux that could potentially be detectable by \textsf{AMS-02}.

In this Letter, we provide a systematic assessment of antinuclei production rates from the decays of beauty baryons and mesons. 
%First, we demonstrate that the predictions of the WL21 tune conflict with hadron spectra and multiplicity measurements at \textsf{LEP} and with the recent \textsf{LHCb} upper limit on ${}^3\overline{\rm He}$ production from $\bar{\Lambda}^0_b$ decays \cite{Moise:2024wqy}. 
We adopt a phenomenologically viable coalescence model from Ref.~\cite{DiMauro:2024kml} that successfully explains the \textsf{ALEPH} measurement of $\overline{\rm D}$ multiplicity \cite{ALEPH:2006qoi} and the \textsf{ALICE} energy spectrum of ${}^3\overline{\rm He}$ \cite{PhysRevC.97.024615}. By constructing a dedicated \textsf{Pythia} tuning that adjusts hadronization, flavor selection parameters, and the unmeasured decay branching ratios of $\bar{\Lambda}^0_b$, we achieve ${}^3\overline{\rm He}$ production rates that comply with the recent \textsf{LHCb} upper limit. Taking all these factors into account, we demonstrate that the enhancement of antinuclei production via $b$-baryon decays is negligible compared to prompt production. 
%This implies that, if the AMS-02 antihelion events are confirmed, an alternative mechanism must be responsible for their origin.

%The $\Lambda^0_b$ is produced with a multiplicity that is a factor of at least 10 larger with respect to the other $b$-baryons. 
%All these particles can travel even meters before decaying and for this reason they are labeled as displaced-vertex decays. 
%nEvents = 1000000
%Sigma_b0  = 4.842e-03 (6.958e-05)
%Sigma_b+  = 4.722e-03 (6.872e-05)
%Sigma_b-  = 4.801e-03 (6.929e-05)
%Xi_b0     = 3.475e-03 (5.895e-05)
%Xi_b-     = 3.690e-03 (6.075e-05)
%Omega_b-  = 4.650e-04 (2.156e-05)
%Lambda_b0 = 4.758e-02 (2.181e-04)
%B0 = 4.245e-01 (6.515e-04)
%B+ = 4.252e-01 (6.521e-04)

%Default CosmiXs:
%antip multiplicity = 5.651e-01 (7.517e-04) (1.784e+00 sigma)
%antiD multiplicity = 8.000e-06 (2.828e-06) (6.163e-01 sigma)
%antiL multiplicity = 8.471e-03 (9.204e-05) (-8.786e-01 sigma)

%Linden setup:
%antip multiplicity = 1.179e+00 (1.086e-03) (1.590e+01 sigma)
%antiD multiplicity = 6.000e-05 (7.746e-06) (6.783e+00 sigma)
%antiL multiplicity = 2.178e-02 (1.476e-04) (7.847e-01 sigma)

%Our tune:
%antip multiplicity = 9.665e-01 (9.831e-04) (1.101e+01 sigma)
%antiD multiplicity = 3.900e-05 (6.245e-06) (5.071e+00 sigma)
%antiL multiplicity = 1.665e-02 (1.290e-04) (1.431e-01 sigma)

\medskip

{\it Current situation --} 
The analysis presented in Ref.~\cite{Winkler:2020ltd} relies on a set of assumptions that may compromise the robustness of its conclusions and therefore merit critical reassessment. Some of the critical points have also been pointed out in Ref.~\cite{Kachelriess:2021vrh}\footnote{The authors of Ref.~\cite{Winkler:2020ltd} have replied to those comments in Ref.~\cite{Winkler:2021cmt}.}.
\begin{itemize}
\item The increase of the \textsf{Pythia} {\tt StringFlav:probQQtoQ} parameter (hereafter {\tt probQQtoQ})\footnote{This parameter models the suppression of diquark production relative to quark production and its default value is 0.081. It affects not only the rates of $\Lambda_b$ but also those of all baryons, such as $p$, $n$, $\Lambda_b^0$, etc. The rate of baryon production in \textsf{Pythia} is directly proportional to the value of this parameter, \textit{i.e.}, a larger value implies a higher baryon production rate.} by a factor of 3 relative to its default value results in an unrealistic overproduction of protons, antiprotons, and other baryons. It even affects important QCD observables at \textsf{LEP}, such as the Thrust and $C$ parameters. In particular, the {\bf WL21 tune} predicts a multiplicity for $\bar{p}+p$ at the $Z$ resonance of 2.36, which is about $40\sigma$ above the updated combination reported by the \textsf{PDG}, $\langle n_{p+\bar{p}} \rangle = 1.050 \pm 0.032$ \cite{ParticleDataGroup:2008zun}. This discrepancy would dramatically increases the predicted rates of antinuclei from the direct production of antinucleons generated in hadronization or resonance decays (also known as the prompt mechanism). In order to compensate for this effect, the authors of Ref.~\cite{Winkler:2021cmt} have significantly decreased the coalescence momentum from a typically used value of 0.18-0.20 GeV \cite{DiMauro:2024kml} to 0.124 GeV. 

\item The {\bf WL21 tune} predicts a multiplicity of $\Lambda_b^0+\bar{\Lambda}_b^0$ (denoted as $\langle n_{\Lambda_b^0} \rangle$) at the $Z$ resonance of 0.044, while the corresponding \textsf{DELPHI} measurement is $0.031 \pm 0.016$ \cite{DELPHI:1996sen}. In contrast, the default \textsf{Monash} tune \cite{Skands:2014pea} of \textsf{Pythia 8} predicts 0.016, which is compatible within $1\sigma$ with the \textsf{LEP} measurement. This suggests that a significant increase of {\tt probQQtoQ} with respect to the \textsf{Pythia 8} default is not required by the $b$-baryon multiplicity measurements.

\item Ref.~\cite{Winkler:2020ltd} used an estimate of the $b$-quark fragmentation function into $b$-baryons of $0.1_{-0.03}^{+0.04}$ taken from a 1998 version of the Review of Particle Physics \cite{ParticleDataGroup:1998hll}, which was mainly tuned to \textsf{LEP} data. A much more recent and precise estimate from the \textsf{HFLAV} collaboration, also based on \textsf{LEP} data \cite{HFLAV:2019otj}, gives 
\[
f(b\rightarrow \Lambda_b) = 0.089 \pm 0.012 .
\]
The {\bf WL21 tune} predicts $f(b\rightarrow \Lambda^0_b) \approx 0.12$, which is about $2.6\sigma$ higher than this recent estimate. Again, an increase by a factor of 3 for {\tt probQQtoQ} does not seem necessary.

\item The model in Ref.~\cite{Winkler:2020ltd} uses the default \textsf{Pythia 8} assumptions for the $\bar{\Lambda}_b^0$ decay modes. This leads to a branching ratio
\[
\operatorname{Br}\left(\bar{\Lambda}_b^0 \rightarrow {}^3\overline{\text{He}} \, X\right) \simeq 1.5 \times 10^{-6},
\]
which is severely excluded by the \textsf{LHCb} collaboration, who measured a 90\% CL upper limit of $6.3 \times 10^{-8}$ for the same process \cite{Moise:2024wqy}.

\end{itemize}

{\it Tuning of the hadronization model} -- 
%Despite the fact that the default \textsf{Pythia 8} predicts a mean multiplicity within $1\sigma$ of the \textsf{DELPHI} data \cite{DELPHI:1996sen}, it yields a prediction that is $3\sigma$ below the mean multiplicity of all $b$-baryons inferred from the measurement of $f(b\rightarrow \Lambda_b)$ reported by the \textsf{HFLAV} collaboration \cite{HFLAV:2019otj}. A naive solution to this issue is to increase the value of {\tt probQQtoQ}, as was done by the authors of Ref.~\cite{Winkler:2020ltd}. Unfortunately, a significant increase or decrease of {\tt probQQtoQ} leads to inconsistent theoretical predictions for several observables, such as baryon spectra, meson spectra, mean multiplicities of identified particles, and certain event shape observables.
%We believe it is not physically justified to compensate for an excessively high baryon multiplicity—resulting from increasing the {\tt probQQtoQ} parameter—by using a much smaller coalescence parameter as done in Ref.~\cite{Winkler:2020ltd}. Instead, we want to first fix the baryon-meson production using LEP data and then tune the coalescence model based on the antinuclei yield. This two-step approach allows us to build a model that is simultaneously compatible with baryon, meson, and antinuclei data.
The default \textsf{Pythia 8} setup reproduces the \textsf{DELPHI} mean multiplicity within $1\sigma$ \cite{DELPHI:1996sen} but predicts a $b$-baryon multiplicity that is $3\sigma$ too low compared to the $f(b\rightarrow \Lambda_b)$ measurement \cite{HFLAV:2019otj}. Although increasing {\tt probQQtoQ} can partly address this, it predicts too high baryon, meson and mean multiplicities. Instead of compensating for this effect with a reduced coalescence parameter, as done in \cite{Winkler:2020ltd}, we propose to first tune \textsf{Pythia 8} to match baryon production measured at \textsf{LEP}, then tune the coalescence model based on antinuclei yields, ensuring compatibility across all datasets.

We therefore perform a comprehensive tune of \textsf{Pythia 8} parameters related directly to both hadronization and flavour selection, using all the relevant measurements from \textsf{LEP}, \textsf{SLD}, the multiplicities of identified particles reported by the \textsf{PDG}, and the fragmentation function reported by \textsf{HFLAV}. The total number of parameters employed in this tuning is 14. For this task, we use \textsf{Pythia 8.311} \cite{Bierlich:2022pfr} to generate Monte Carlo (MC) samples, \textsf{Rivet 3.1.6} \cite{Bierlich:2019rhm} for the experimental analyses at the particle level, and \textsf{Professor 2.4.0} \cite{Buckley:2009bj} for the tuning of the parameters. The total number of measurement bins is 4185. The tuning setup is similar to that used in earlier studies (see, e.g., Refs.~\cite{Amoroso:2018qga,Jueid:2022qjg,Jueid:2023vrb,Arina:2023eic}). Technical details of this tune can be found in Appendix \ref{app:tuning}. The fitting procedure yields a value of {\tt probQQtoQ} equal to 0.111. The corresponding goodness-of-fit per degree of freedom at the minimum is $\chi^2/N_{\rm df} = 4943.16/4171 \approx 1.18$. We also estimate the uncertainties on the tuned parameters using the eigentunes method, which is based on the diagonalization of the Hessian matrix near the minimum (details can be found in Ref.~\cite{Jueid:2023vrb}). The comparison of our results for the mean multiplicities of $\Lambda$, $\Lambda^0_b$, $p/\bar{p}$, and $\pi^\pm$ with experimental measurements is shown in Fig.~\ref{fig:comparison}. We observe that the best-fit value for $f(b\rightarrow \Lambda_b)$ obtained from our tune is about $2\sigma$ smaller than the \textsf{HFLAV} estimate, while all the other predictions are fully compatible with the \textsf{LEP} data. The difference is only at $1\sigma$ if we consider also the error on $f(b\rightarrow \Lambda_b)$ coming from the fit we perform (orange band in Fig.~\ref{fig:comparison}). For the multiplicity of $\Lambda_b^0+\bar{\Lambda}_b^0$, we obtain 0.021, which is fully compatible with the \textsf{DELPHI} measurement, $0.031 \pm 0.016$ \cite{DELPHI:1996sen}. For the remainder of the Letter, this tune will be referred to as {\bf Had.tune}.

\begin{figure}[!t]
\centering
\includegraphics[width=0.49\linewidth]{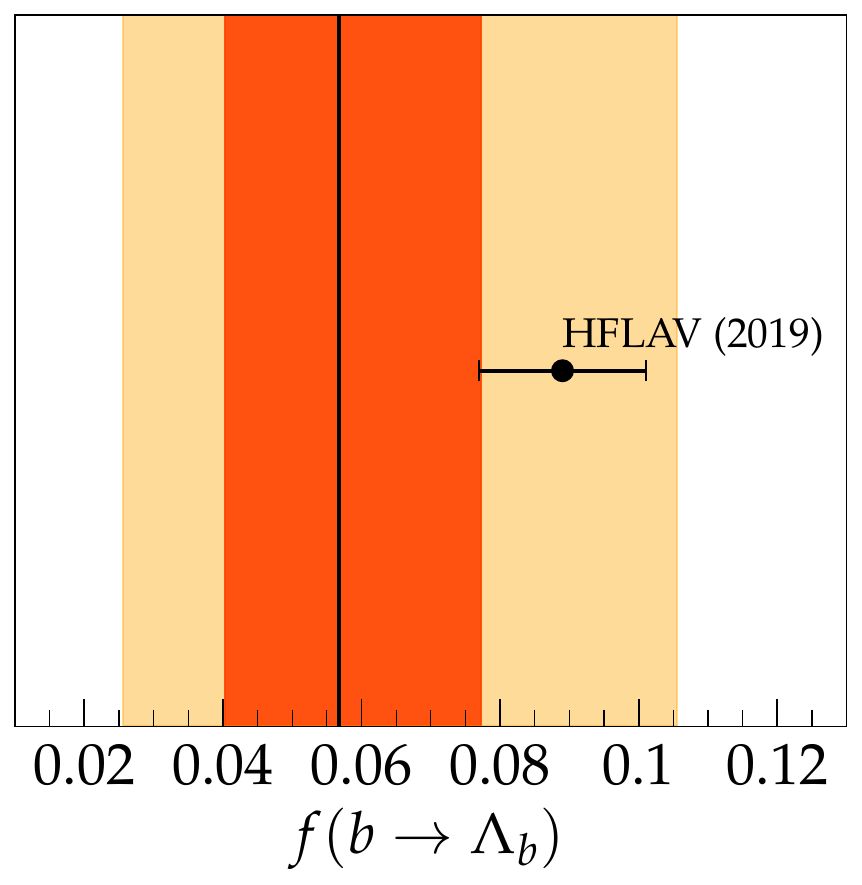}
\includegraphics[width=0.49\linewidth]{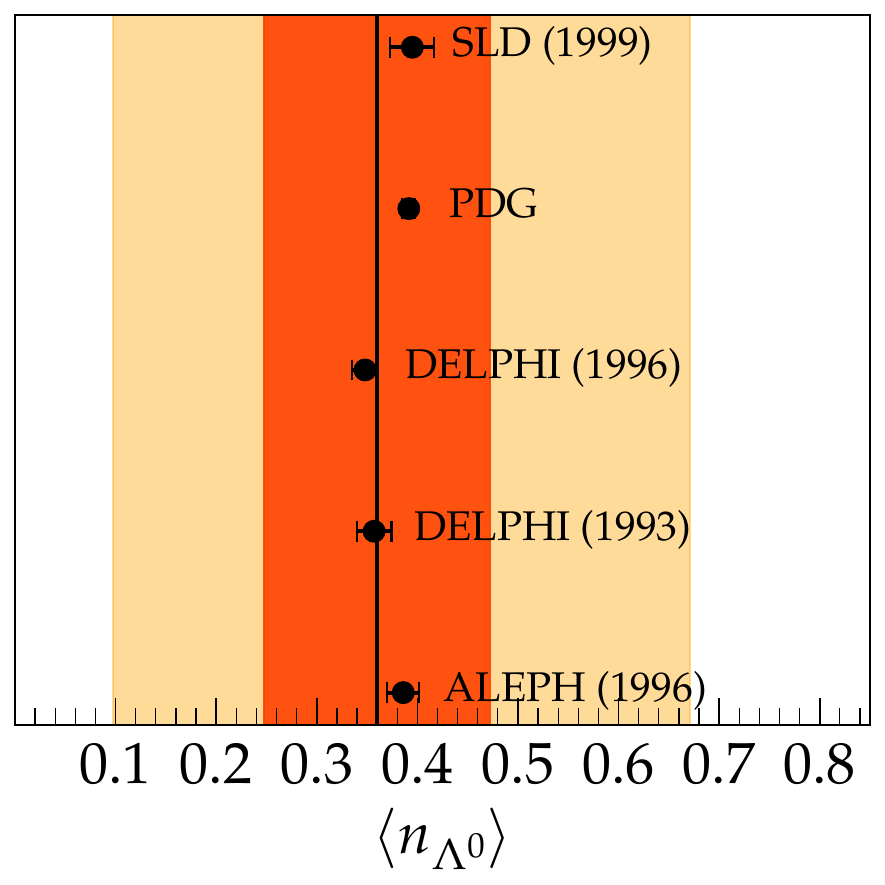}
\includegraphics[width=0.49\linewidth]{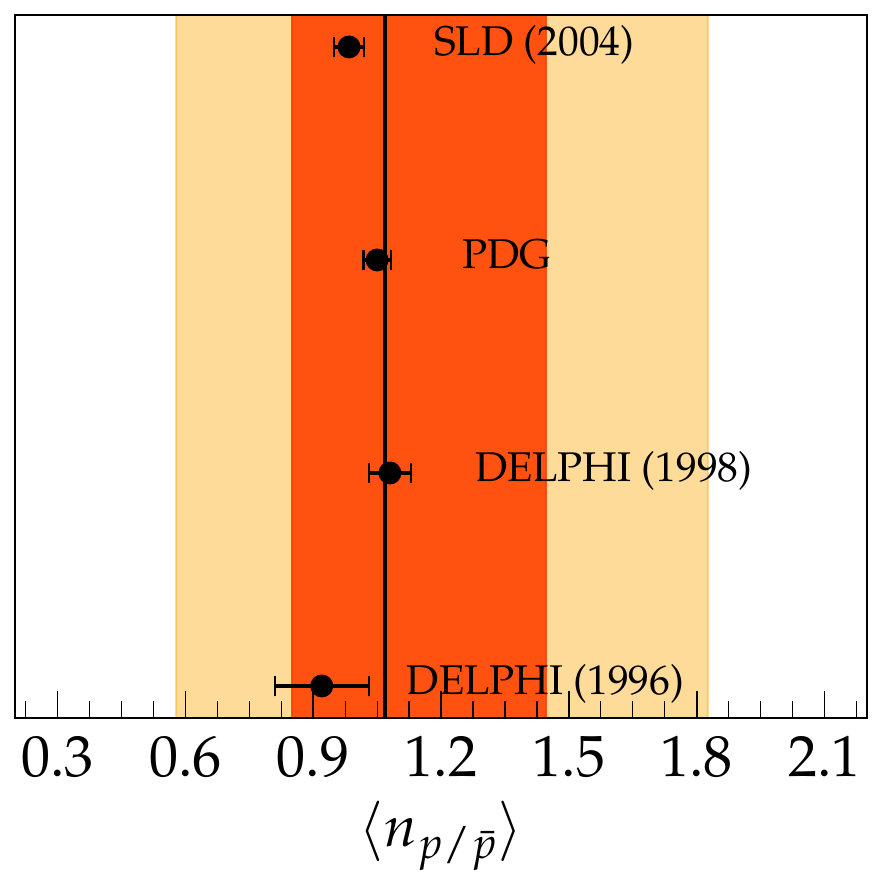}
\includegraphics[width=0.49\linewidth]{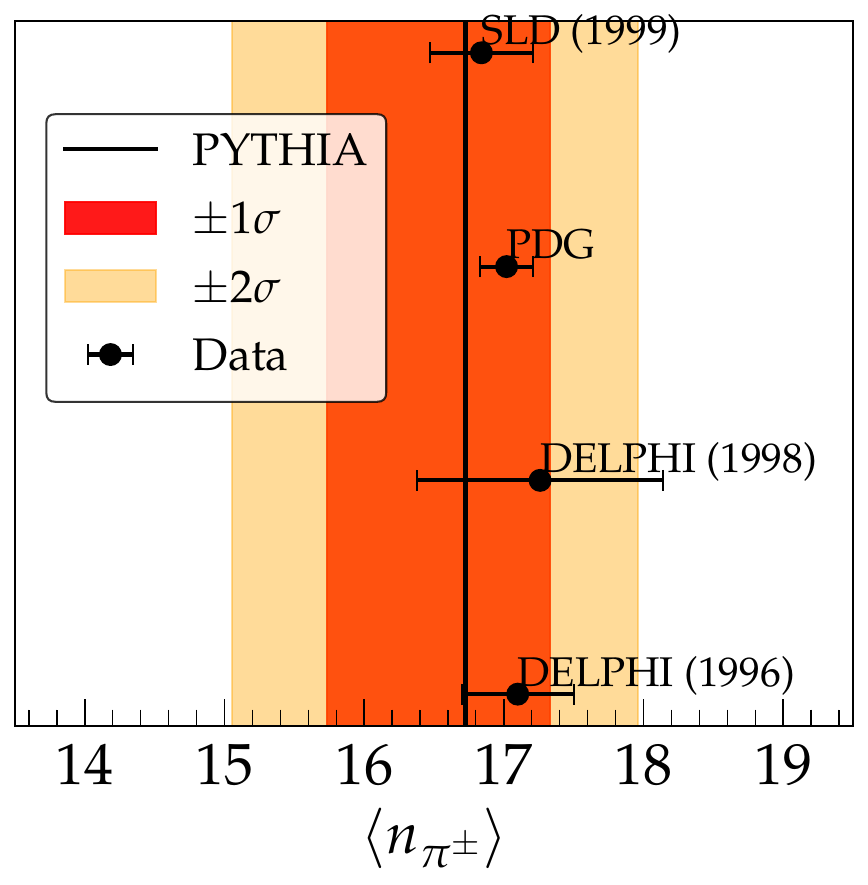}
\caption{Comparison between the theory predictions obtained at the best-fit parameter point of \textsf{Pythia 8} with the {\bf Had.tune} and the experimental values for $f(b \to \Lambda_b)$ (upper left), the multiplicity of $\Lambda^0$ (upper right), the multiplicity of $p/\bar{p}$ (lower left), and the multiplicity of $\pi^\pm$ (lower right). The $1\sigma$ and $2\sigma$ theory uncertainties are shown as red and orange bands, respectively. Here $\Lambda_b$ refers to all the $b$-baryons and not only $\Lambda_b^0$. Data are taken from Refs. \cite{ALEPH:1996oqp,DELPHI:1993vpj,DELPHI:1996sen,DELPHI:1998cgx,HFLAV:2019otj,ParticleDataGroup:2008zun,SLD:1998coh,SLD:2003ogn}.}
\label{fig:comparison}
\end{figure}

\medskip

{\it Coalescence model implemented in \textsf{Pythia} --} In Ref.~\cite{DiMauro:2024kml}, some of the authors of this paper implemented four different coalescence models that consistently reproduce the results of \textsf{ALEPH} for the $\overline{\text{D}}$ multiplicity in $Z$-resonance hadronic decays \cite{2006192}. Two of these models are very simple, taking into account either only the difference in momentum or both the difference in momentum and the spatial separation between antinucleons. The other two models are more sophisticated, employing the Wigner formalism with either a Gaussian or Argonne wave function to account for possible antinucleon correlations in space and momentum. The model we use, inspired by Ref.~\cite{DiMauro:2024kml}, is a fully MC-based approach implemented in \textsf{Pythia 8.311}: we generate DM particle annihilation events, then search for every pair of $\bar{n}$ and $\bar{p}$ produced in the annihilation process, and decide whether a $\overline{\rm D}$ is formed. Employing an MC generator enables us to properly take into account both spatial and momentum correlations between antinucleons (see Ref.~\cite{DiMauro:2024kml} for further details).

As a proof of principle, we adopt in this Letter a simple coalescence model that imposes criteria on the momentum difference ($\Delta p < p_c$) and the spatial separation ($\Delta r$) in the center-of-mass frame of the $\bar{p}$-$\bar{n}$ system\footnote{Ref.~\cite{DiMauro:2024kml} shows that very similar $\overline{\text{D}}$ spectra are obtained when using different coalescence models that are properly tuned on \textsf{ALEPH} data \cite{ALEPH:2006qoi}. Therefore, the conclusions of this Letter will not change if we consider the Wigner formalism.}. Including the criterion related to the $\bar{n}$-$\bar{p}$ distance is particularly important for antinuclei generated from off-vertex particle decays.

In Ref.~\cite{DiMauro:2024kml} we calibrated the coalescence model on the \textsf{ALEPH} $\overline{\text{D}}$ data and found that, by fixing $\Delta r < 3$\,fm\footnote{This value is motivated by the typical sizes of the $D$ and He nuclei.}, a coalescence momentum of $p_c = 0.21 \pm 0.02$\,GeV is needed. Here, instead, we fix the coalescence model for the production rate of ${}^3\overline{\rm He}$ using the measurement by \textsf{ALICE} in $pp$ collisions at $\sqrt{s} = 7$ TeV \cite{PhysRevC.97.024615}. In fact, no measurements for ${}^3\overline{\rm He}$ are available from \textsf{LEP}. Since the production of baryons and antinuclei in $pp$ collisions is expected to differ from that at \textsf{LEP}, we first tune \textsf{Pythia} to correctly predict the proton yield measured by \textsf{ALICE}. Details of the model setup are provided in Appendix \ref{appx:ALICEcalib}. The predicted proton multiplicity is $1\sigma$ compatible with the \textsf{ALICE} measurement \cite{PhysRevC.97.024615}. Once the nucleon spectra are calibrated, we fit the \textsf{ALICE} helion yield and find that, for $\Delta r < 3$\,fm, the best-fit coalescence momentum is $p_c = 0.20 \pm 0.01$\,GeV. Note that this value is compatible within $1\sigma$ with the one obtained from the \textsf{ALEPH} data for $\overline{\text{D}}$ \cite{DiMauro:2024kml}. In Appendix \ref{appx:ALICEcalib} we show a comparison of the $p+\bar{p}$, $\overline{\text{D}}$, and ${}^3\overline{\text{He}}$ spectra obtained with our model and the corresponding \textsf{ALICE} data. The value obtained from the fit to the ${}^3\overline{\text{He}}$ spectra will be used in the remainder of this Letter.

\medskip

{\it $b$-baryon and $B$-meson fragmentation functions--} 
As we will show in the results section, the production of antideuterons from weakly decaying $B$-mesons is roughly as important as that from $b$-baryons. $B$-mesons—namely, $B^0$ and $B^{\pm}$ particles—are produced with a multiplicity that is two to three times higher than that of $b$-baryons (see, e.g., Ref.~\cite{LHCb:2023wbo}). Therefore, we take into account not only the $b$-baryon fragmentation function, $f(b\rightarrow \Lambda_b)$, but also the $B$-meson fragmentation functions, $f(b\rightarrow B^0, B^0_s)$, as reported by the \textsf{HFLAV} collaboration \cite{HFLAV:2019otj}.

When comparing the {\bf Had.tune} model predictions for the fragmentation functions into weakly decaying $b$-hadrons with the estimates from \textsf{HFLAV} \cite{HFLAV:2019otj}, we find a noticeable discrepancy. This discrepancy can lead to an imprecise prediction for the $\overline{\text{D}}/{}^3\overline{\text{He}}$ spectra from weakly decaying $b$-hadrons. Therefore, we refine the value of {\tt probQQtoQ} to best match the measured $f(b\rightarrow \Lambda_b, B^0, B^0_s)$ specifically for antinuclei generated from $b$-hadrons\footnote{For the prompt production, we retain the {\bf Had.tune}, which yields the correct hadron yields.}. To do this, we perform a simple $\chi^2$ analysis comparing the predictions from {\bf Had.tune} for the fragmentation functions with the \textsf{HFLAV} estimates. In the {\bf Had.tune} model, we fix all parameters except for {\tt probQQtoQ} and obtain a best-fit value of $0.19\pm0.03$. Thanks to this refinement, our model becomes compatible with all $b$-hadron fragmentation functions within the $1\sigma$ uncertainties (see Fig.~\ref{fig:frag}). In the rest of the Letter, we refer to this tuning as {\bf Had.tune+QQ}.

\begin{figure}
\centering
\includegraphics[width=0.99\linewidth]{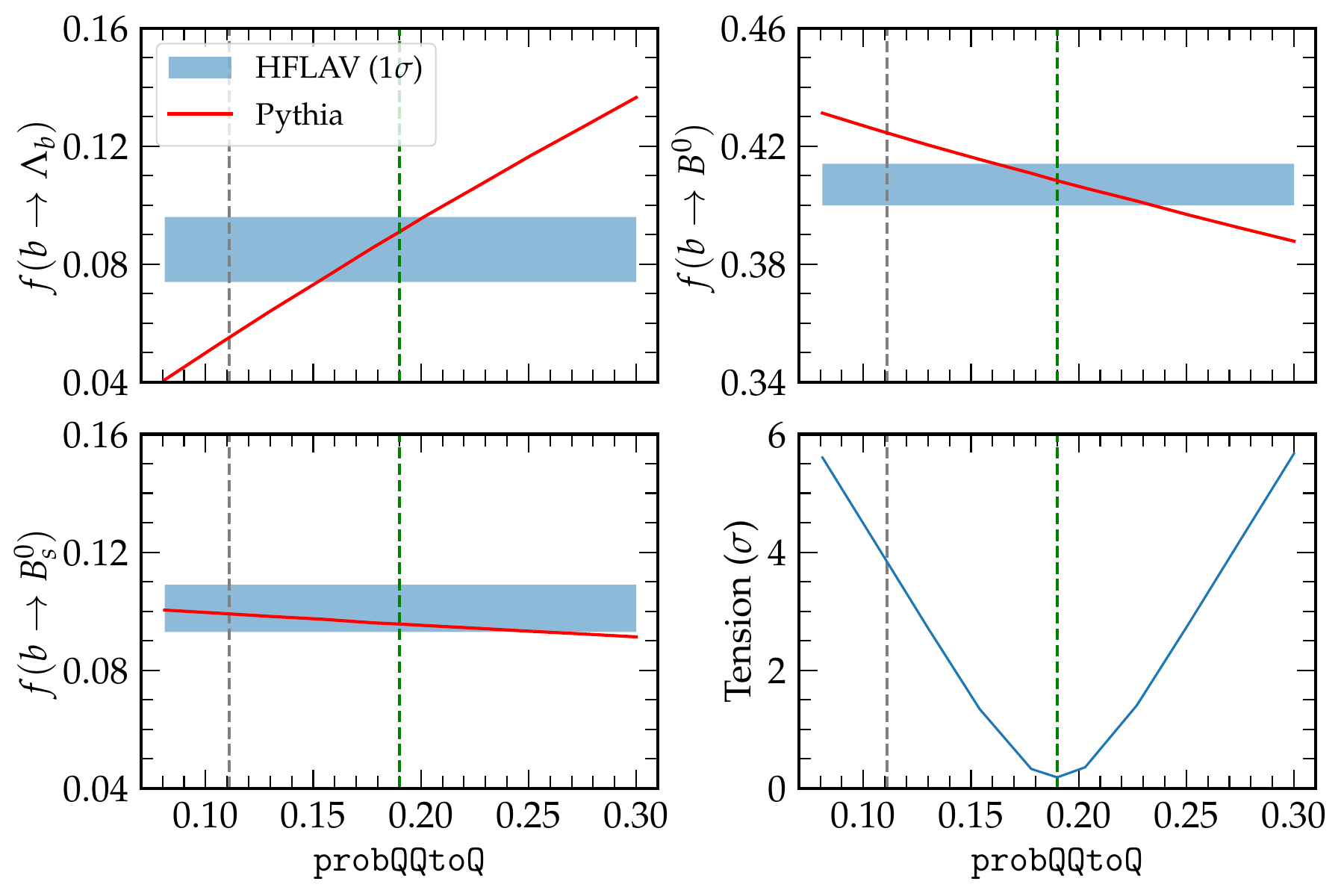}
    \caption{Comparison between the fragmentation functions for $\Lambda^0_b$ (top left panel), $B^0$ (top right panel) and $B^0_s$ (bottom left panel) obtained with \textsf{Pythia} (red line) and those reported by \textsf{HFLAV} \cite{HFLAV:2019otj} (blue band, $1\sigma$ error) as a function of {\tt probQQtoQ}. The bottom right panel shows the overall tension between \textsf{Pythia} and \textsf{HFLAV} (combining all considered fragmentation functions) with respect to {\tt probQQtoQ}. The best-fit value of {\tt probQQtoQ} from our hadronization tune is indicated by a gray dashed line, and the value that best fits the fragmentation functions is shown as a green dashed line.}
    \label{fig:frag}
\end{figure}

\medskip

\begin{figure*}[ht!]
\includegraphics[width=0.49\linewidth]{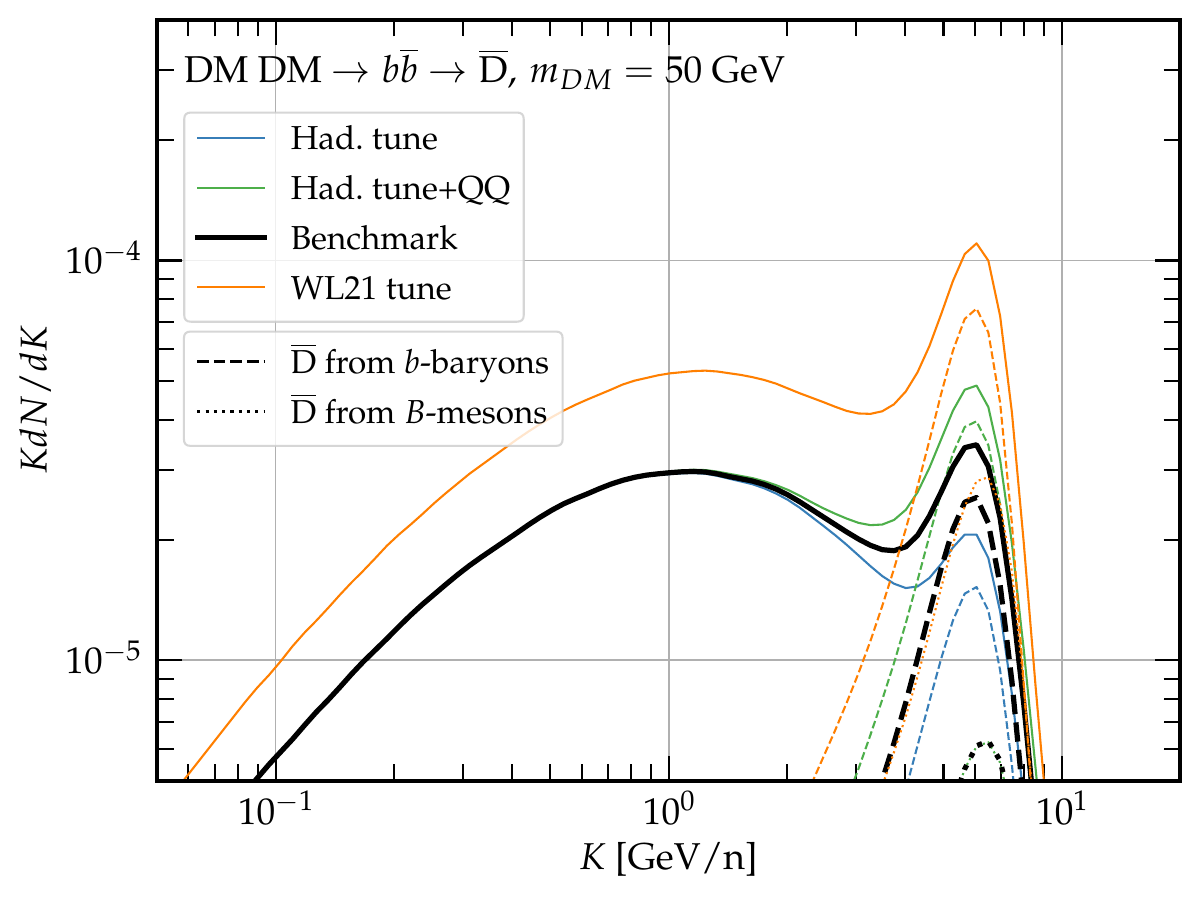}
\includegraphics[width=0.49\linewidth]{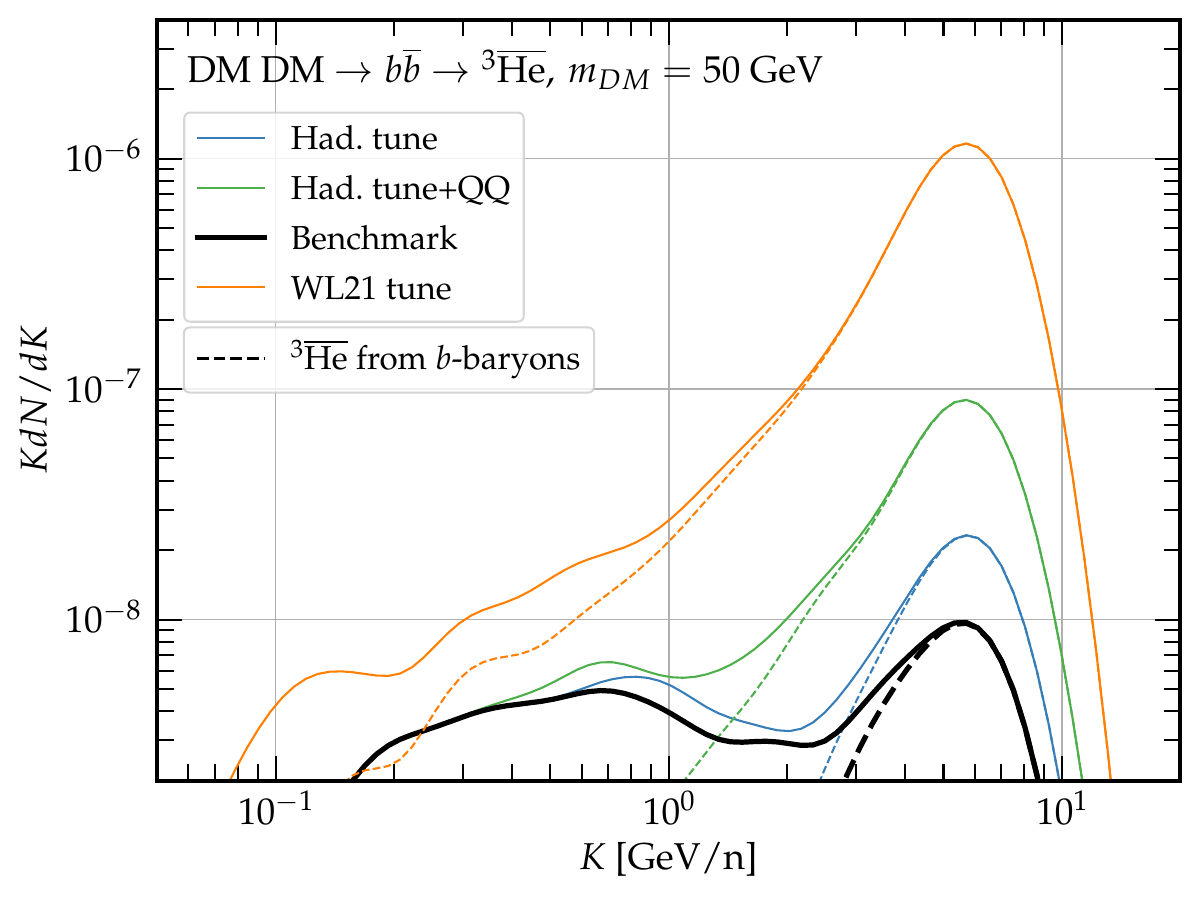}
    \caption{Source spectra for the production of $\overline{\text{D}}$ (left panel) and ${}^3\overline{\text{He}}$ (right panel) from DM annihilations into $b\bar{b}$ with a mass of $50$ GeV. For the prompt production of $\overline{\text{D}}/{}^3\overline{\text{He}}$, we always use \textbf{Had.tune}. For the $\overline{\text{D}}/{}^3\overline{\text{He}}$ produced by weakly decaying $b$-hadrons (and $B$-mesons), represented by dashed (dotted) lines, we show the results for different cases: {\bf Had.tune} (blue), {\bf Had.tune+QQ} (green), {\bf Benchmark} (black), and {\bf WL21 tune} (orange). Solid lines represent the sum of the prompt production, weakly decaying $b$-baryon, and $B$-meson contributions.}
    \label{fig:spectra}
\end{figure*}

{\it Tuning the decay branching ratios of $\bar{\Lambda}^0_b$--} 
The models {\bf Had.tune} and {\bf Had.tune+QQ} predict branching ratios for $\operatorname{BR}\left(\bar{\Lambda}_b^0 \rightarrow {}^3\overline{\text{He}}\, X\right)$ of $(2.2\pm0.3)\times10^{-7}$ and $(5.6\pm0.4)\times10^{-7}$, respectively—about factors of 3 and 8 above the upper limit of $6.8\times10^{-8}$ reported by \textsf{LHCb} \cite{Moise:2024wqy}. These differences are less dramatic than those obtained in Ref.~\cite{Winkler:2020ltd}, which reports $\operatorname{BR}\left(\bar{\Lambda}_b^0 \rightarrow {}^3\overline{\mathrm{He}}\, X\right) \simeq 1.5 \times 10^{-6}$. %\textcolor{red}{WE SHOULD CHANGE THIS When testing the same {\tt probQQtoQ} value as in {\bf WL21} (three times the default) but using our coalescence model, we find 
%\[
%\operatorname{BR}\left(\bar{\Lambda}_b^0 \rightarrow {}^3\overline{\mathrm{He}}\, X\right) = (8.9 \pm 0.1) \times 10^{-6}.
%\]}

The prediction of this branching ratio depends mainly on the coalescence model, the {\tt probQQtoQ} value and the decay modes of the $\Lambda_b^0$. The former two ingredients have been fixed already. Therefore, to improve our model and ensure consistency with the \textsf{LHCb} upper limit, we modify the tabulated branching ratios of $\Lambda_b^0$ into di-quark modes in \textsf{Pythia}. In particular, \textsf{Pythia} assigns the main decay channel to a mode involving three quarks plus one di-quark state. The channel with the highest branching ratio is 
\begin{equation}
\label{eq:Br1}
\operatorname{BR} \left (\Lambda_b^0 \rightarrow \bar{u} d c \, (ud)_0\right),
\end{equation}
fixed at approximately $53\%$. In contrast, the channel most relevant for antinuclei production is 
\begin{equation}
\label{eq:Br2}
\operatorname{BR}\left(\Lambda_b^0 \rightarrow \bar{u} d u \, (ud)_0\right),
\end{equation}
which occurs at around $1.2\%$. This difference is largely related to the magnitudes of the Cabibbo–Kobayashi–Maskawa matrix elements, $\left|V_{ub}\right|$ and $\left|V_{cb}\right|$, which control the transitions $\bar{b}\rightarrow \bar{u}$ and $\bar{b}\rightarrow \bar{c}$, respectively. In particular, $\left|V_{ub}\right|^2/\left|V_{cb}\right|^2\approx 0.01$, about a factor of 2 smaller than the ratio $1.2/53\approx 0.022$ implemented in \textsf{Pythia}. 
This difference is compensated by the fact that the allowed phase space prefers a higher multiplicity in $\bar{b} \rightarrow \bar{u}$ than in $\bar{b} \rightarrow \bar{c}$.  The branching ratios in Eq.~\ref{eq:Br1} and Eq.~\ref{eq:Br2} are however not measured by any experiment yet but can be {\it thought} as external input parameters in \textsf{Pythia} whose values are just guessed and thus subject to large theoretical uncertainties.

The second most important decay channel involves the production of leptons and charmed baryons via $\bar{\Lambda}_b^0 \rightarrow \Lambda_c^-\ell^+ \nu_\ell$.
This decay is measured to occur with a branching fraction of $(6.2^{+1.4}_{-1.3})\%$ \cite{ParticleDataGroup:2024cfk}, while \textsf{Pythia} predicts a value of $5.96\%$, which is compatible. 

The main change we applied in \textsf{Pythia} is a reduction of the branching ratio $\operatorname{BR}\left(\Lambda_b^0 \rightarrow \bar{u} d u \, (ud)_0\right)$
to $10^{-3}$. With this adjustment, we obtain 
\[
\operatorname{BR}\left(\bar{\Lambda}_b^0 \rightarrow {}^3\overline{\text{He}}\, X\right) = (7\pm 1)\times 10^{-8},
\]
which is consistent with the \textsf{LHCb} upper limit. In Appendix \ref{app:Brlambdab}, we detail the modifications applied to the $\Lambda_b^0$ branching ratio values and compare the main decay channels with experimental measurements. Overall, our model yields predictions for the main decay modes of $\Lambda_b^0$ that are compatible with observations. In the rest of the paper, we label this setup as {\bf Benchmark}.

\medskip

{\it Results for the spectra of $\overline{\rm D}$ and ${}^3\overline{\rm He}$--} 
We now present our predictions for the antideuteron and antihelion spectra. We choose a DM mass of 50 GeV with the $b\bar{b}$ annihilation channel
%\footnote{The choice of a 50 GeV DM mass is motivated by the WIMP scenario and by the fact that $\bar{\Lambda}^0_b$ production of ${}^3\overline{\rm He}$ is maximized at relatively low masses.}. 
We run simulations with $10^{10}$ annihilation events, which yield, for the \textbf{Benchmark}, a total of approximately $2\times10^6$ $\overline{\text{D}}$ and 500 ${}^3\overline{\text{He}}$ events. Fig.~\ref{fig:spectra} displays the resulting spectra: the total spectrum (solid line) along with the separate contributions from $b$-baryons (dashed line) and $B$-mesons (dotted line).

Under the \textbf{Benchmark}, $B$-mesons contribute about $20\%$ of the total $b$-hadron yield in the $\overline{\text{D}}$ spectrum, while they do not contribute to the ${}^3\overline{\text{He}}$ spectrum. Overall, antinuclei produced from $b$-hadron decays are a minor component, contributing only at the high-energy end of the spectrum (i.e., for energies close to the kinematic cutoff of the DM mass), with contributions of about $8\%$ and $16\%$ of the total for $\overline{\text{D}}$ and ${}^3\overline{\text{He}}$, respectively. Even when focusing solely on the highest energies, $b$-hadrons account for at most about $50\%$ of the total counts.

The use of the default \textsf{Pythia} settings for the $\Lambda_b^0$ decay channels, as done in Ref.~\cite{Winkler:2020ltd}, has important implications. In the case labeled as \textbf{Had.tune+QQ}, where the default $\Lambda_b^0$ decay modes are employed, the spectra are approximately $30\%$ larger for $\overline{\text{D}}$ and a factor of 10 larger for ${}^3\overline{\text{He}}$ compared to the \textbf{Benchmark}. Thus, the modifications we introduced to the rare $\bar{\Lambda}_b^0$ decay modes—made to conform with the \textsf{LHCb} upper limits on $\operatorname{BR}\left(\bar{\Lambda}_b^0 \rightarrow {}^3\overline{\text{He}}\, X\right)$—have a significant impact on the ${}^3\overline{\text{He}}$ spectrum. In contrast, employing the {\tt probQQtoQ} value from \textbf{Had.tune}, which is closer to the default \textsf{Pythia} setting, leads to a reduced $b$-hadron contribution; in particular, the ${}^3\overline{\text{He}}$ spectrum is reduced by a factor of 3.

We also demonstrate that we recover the significant enhancement reported in Ref.~\cite{Winkler:2020ltd} when using a {\tt probQQtoQ} value three times larger than the default, the default $\bar{\Lambda}_b^0$ branching ratios and $p_c=0.124$ GeV. In that scenario, the enhancement in the $\overline{\text{D}}$ and ${}^3\overline{\text{He}}$ spectra due to the weakly decaying $b$-hadrons is approximately a factor of 3 and 50, respectively, relative to the predictions obtained with the default \textsf{Pythia} setting. However, as noted above, this model is severely excluded by the \textsf{LHCb} upper limits.

\medskip

{\it Conclusions} -- In this Letter we have developed a hadronization model for the prompt production of antinucleons that is fully compatible with a comprehensive \textsf{LEP} dataset. In addition, we tuned the antinucleon yield from weakly decaying $b$-hadrons using multiplicity data and $b$-hadron fragmentation function estimates. We constrained the coalescence process by fitting \textsf{ALICE} data for the ${}^3\overline{\text{He}}$ spectrum and found the model to be compatible at $1\sigma$ with the \textsf{ALEPH} $\overline{\text{D}}$ data. Finally, we adjusted the \textsf{Pythia} settings for the decay of $\bar{\Lambda}^0_b$ to be consistent with the \textsf{LHCb} upper limit for $\operatorname{BR}\left(\bar{\Lambda}^0_b \rightarrow {}^3\overline{\text{He}}\, X\right)$. With these refinements, we provide the most precise estimates to date for the production of $\overline{\text{D}}$ and ${}^3\overline{\text{He}}$ from $B$-meson and $b$-baryon weak decays, finding that their contribution is negligible compared to that from prompt production. This implies that if the tentative detections of antihelion events by \textsf{AMS-02} are confirmed, new production mechanisms must be considered.

\medskip

\begin{acknowledgments}
The authors thank Francesca Bellini, Fiorenza Donato, Nicolao Fornengo, Tim Linden, and Martin Winkler for providing very helpful comments on the manuscript and the analysis.

M.D.M. and J.K.~acknowledge support from the research grant {\sl TAsP (Theoretical Astroparticle Physics)} funded by Istituto Nazionale di Fisica Nucleare (INFN) and the PRIN project N.20222BBYB9 EXSKALIBUR – Euclid-Cross-SKA: ``Likelihood Inference Building for Universe's Research'' and the funding provided by the INFN Torino Section. 
The work of A.J. is supported by the Institute for Basic Science (IBS) under the project code, IBS-R018-D1. R.R.dA. acknowledges support from the Ministerio de Ciencia y Innovación (PID2020-113644GB-I00) and the GVA Research Project {\sl Sabor y Origen de la Materia (SOM)} (PROMETEO/2022/069). 
\end{acknowledgments}

%\newpage

%\appendix

\bibliographystyle{apsrev4-1}
\bibliography{main.bib}

%\clearpage
%\maketitle
%\onecolumngrid
%\begin{center}
%{\bf \large SUPPLEMENTAL MATERIAL}
%{Mattia Di Mauro, Nicolao Fornengo, Adil Jueid, Roberto Ruiz de Austri, Chiara Arina, and Francesca Bellini}
%\vspace{0.05in}
%\end{center}
%\onecolumngrid
%%%%%%%%%% Merge with Supplemental material %%%%%%%%%%
%\setcounter{equation}{0}
%\setcounter{figure}{0}
%\setcounter{table}{0}
%\setcounter{section}{0}
%\setcounter{page}{1}
%\makeatletter
%\renewcommand{\theequation}{S\arabic{equation}}
%\renewcommand{\thefigure}{S\arabic{figure}}

%\section{Details about the tuning of \textsf{PYTHIA} parameters}
%\label{app:wigner}

\clearpage
\maketitle
\onecolumngrid
\begin{center}
%\textbf{\large CosmiXs: spectra of cosmic antideuterons from dark matter interactions} \\ 
%\vspace{0.05in}
{\bf \large SUPPLEMENTAL MATERIAL}
%{Mattia Di Mauro, Nicolao Fornengo, Adil Jueid, Roberto Ruiz de Austri, Chiara Arina, and Francesca Bellini}
\vspace{0.05in}
\end{center}
\onecolumngrid
%%%%%%%%%% Merge with Supplemental material %%%%%%%%%%
\setcounter{equation}{0}
\setcounter{figure}{0}
\setcounter{table}{0}
\setcounter{section}{0}
\setcounter{page}{1}
\makeatletter
\renewcommand{\theequation}{S\arabic{equation}}
\renewcommand{\thefigure}{S\arabic{figure}}

\section{Tuning of the hadronization model }
\label{app:tuning}

\begin{table*}[!htb]
\setlength\tabcolsep{13pt}
  \begin{center}
    \begin{tabular}{lccc}
      \toprule \toprule
      Parameter & Range & \textsf{Monash} & Had.~tune \\ 
      \toprule
      \verb|StringZ:aLund| & $0.0$ -- $2.0$ & $0.68$ & $0.7832 \pm 0.0123$ \\ 
      \verb|StringZ:bLund| & $0.2$ -- $2.0$ & $0.98$ & $1.1729 \pm 0.0100$\\ 
      \verb|StringZ:aExtraDiquark| & $0.0$ -- $2.0$ & $0.97$ & $0.9251 \pm 0.0175$ \\   
      \verb|StringFlav:ProbStoUD| & $0.0$ -- $1.0$ & $0.217$ & $0.2265 \pm  0.0016$ \\
      \verb|StringFlav:mesonUDvector| & $0.0$ -- $3.0$ & $0.50$ & $0.6655 \pm 0.0152$  \\
      \verb|StringFlav:mesonSvector| & $0.0$ -- $3.0$ & $0.55$ & $0.5842 \pm 0.0177$ \\
      \verb|StringFlav:etaSup| & $0.0$ -- $1.0$ & $0.60$ & $0.6499 \pm 0.0005$ \\
      \verb|StringFlav:etaPrimeSup| & $0.0$ -- $1.0$ & $0.12$ & $0.1778 \pm 0.0037$ \\
      \verb|StringFlav:probQQtoQ| & $0.0$ -- $1.0$ & $0.081$ & $0.1112 \pm 0.0008$ \\
      \verb|StringFlav:probSQtoQQ| & $0.0$ -- $1.0$ & $0.915$ & $0.9791 \pm 0.0061$ \\
      \verb|StringFlav:probQQ1toQQ0| & $0.0$ -- $1.0$ & $0.0275$ & $0.8761 \pm 0.0171$ \\
      \verb|StringFlav:popcornSpair| & $0.0$ -- $1.0$ & $0.50$ & $0.6108 \pm 0.0241$ \\
      \verb|StringFlav:popcornSmeson| & $0.0$ -- $1.0$ & $0.90$ & $0.8306 \pm 0.0231$\\
      \verb|StringFlav:popcornRate| & $0.0$ -- $1.0$ & $0.50$ & $0.4117 \pm 0.0055$ \\
     \bottomrule \bottomrule
    \end{tabular}
  \end{center}
    \caption{\label{tab:tune} This table shows the \textsf{Pythia 8} parameters we have included in the tune of this paper, which can be split into hadronization function parameters (those starting with {\tt StringZ:}) and flavor selection parameters (those starting with {\tt StringFlav:}). We show their allowed range in \textsf{Pythia 8}, their default values obtained in the \textsf{Monash} tune and the results of this study along with their \textsf{MIGRAD} errors.} 
\end{table*}

\begin{figure*}[!t]
    \centering
    \includegraphics[width=0.8\linewidth]{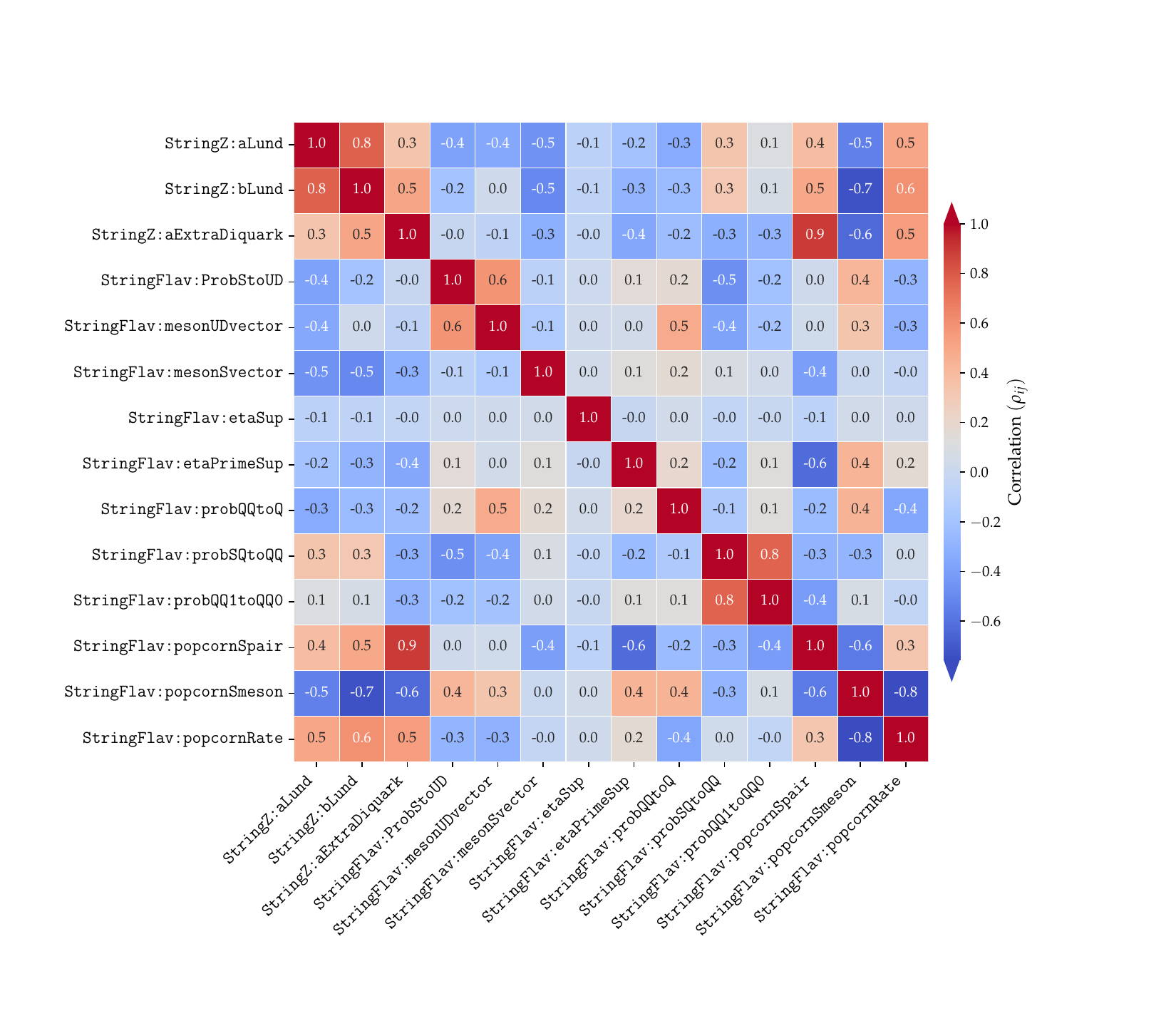}
    \vspace{-1cm}
    \caption{Correlation matrix among the parameters of \textsf{Pythia} at the minimum.}
    \label{fig:tune:correlation}
\end{figure*}

In this appendix we provide details of the tuning setup and show some results regarding the values of the parameters at the minimum. We close this section with a discussion of the theoretical uncertainties on the parameters we derive in this work. We perform a comprehensive tune of \textsf{Pythia 8} parameters directly related to both the hadronization as well as the flavour selection using all the relevant measurements at \textsf{LEP} and \textsf{SLD}. For this task, we use \textsf{Pythia 8.311} \cite{Bierlich:2022pfr} to generate Monte Carlo (MC) samples while \textsf{Rivet 3.1.6} is used for the experimental analyses at the particle level \cite{Bierlich:2019rhm} and \textsf{Professor 2.4.0} is used for the tuning of the parameters \cite{Buckley:2009bj}. The \textsf{Professor} toolkit is a based on a method which permits the optimization of all the parameters using analytical expressions that are cast as polynomials whose coefficients are determined by fitting MC simulated predictions generated for a set of randomly selected parameter points.  Then, the optimal values of the parameters are then obtained with the help of a standard $\chi^2$ minimization using \textsf{Minuit} \cite{James:1975dr}. In this study, we tune 14 parameters listed in Tab.~\ref{tab:tune} and thus we generated MC samples for 6000 random points in the parameter space. The best-fit parameters are determined by minimizing the following $\chi^2$ 
\begin{equation}
 \chi^2 = \sum_{i,j}  \bigg({\cal O}_i - {\cal O}_i^{\rm exp}\bigg)\frac{1}{\sigma_{ij}^2} \bigg({\cal O}_j - {\cal O}_j^{\rm exp}\bigg),
\label{eq:GoF}
\end{equation}
with ${\cal O}_i$ being the MC prediction for the observable $i$ which is cast here as a third order polynomial in the parameters ${\bf x} = \{x_1, \cdots, x_{14} \}$; ${\cal O}_i \equiv \sum_{i+j+k\leq 3} c_{ijk} x^i_\alpha x^j_\beta x^k_\gamma$, and the sum is over all the measurements being included. In equation \ref{eq:GoF}, $\sigma_{ij}$ represents the covariance matrix which is assumed to be diagonal, {\it i.e.} $\sigma_{ij} = \sigma_i \delta_{ij}$ where $\sigma_i$ being the total error which is defined as the sum in quadrature of the experimental error, statistical error due to the limited size of the simulated MC event samples and a flat $5\%$ theory uncertainty which is added to avoid overfitting effects and model the limited accuracy in both the perturbative and non-perturbative regimes:
\begin{eqnarray}
\sigma_i \equiv \sqrt{\sigma_{i, \rm exp}^2 + \sigma_{i, {\rm MC}}^2 + [0.05 \times {\cal O}_i]^2}.
\end{eqnarray}
A good fit is achieved when the goodness-of-fit per number of degrees-of-freedom ($\chi^2/N_{\rm df}$) is of order ${\cal O}(1)$ where $N_{\rm df}$ is defined as the number of measurement data points minus the number of parameters, $N_{\rm df} = \sum_{i} {\cal O}_i - N_{\rm params}$. To achieve a good model we include all the possible measurements of event shapes, charged multiplicities, spectra of baryons and mesons, multiplicities of identified particles including vector mesons, and angular correlations between $\Lambda$ baryons. In this work, we use measurements performed at the $Z$-pole by \textsf{ALEPH} \cite{ALEPH:1991ldi,ALEPH:1995aqx,ALEPH:1996oqp,ALEPH:1996pxg,ALEPH:1999udi,ALEPH:2001pfo,ALEPH:2003obs}, \textsf{DELPHI} \cite{DELPHI:1990ohs,DELPHI:1993vpj,DELPHI:1994qgk,DELPHI:1994aml,DELPHI:1995kfu,DELPHI:1995ase,DELPHI:1996xro,DELPHI:1996sen,DELPHI:1996ztb,DELPHI:1998cgx,DELPHI:1999hkl,DELPHI:2000uri}, \textsf{HFLAV} \cite{HFLAV:2019otj},  \textsf{L3} \cite{L3:1991nwl,L3:1992nwf,L3:1994gkb,L3:1994jps,L3:2004cdh}, \textsf{OPAL} \cite{OPAL:1991lzj,OPAL:1994zan,OPAL:1995ebr,OPAL:1996ecm,OPAL:1995tjq,OPAL:1996gsw,OPAL:1998wmk,OPAL:1998enc,OPAL:1998arz,OPAL:1998tlp,OPAL:2000dkf,OPAL:2000rtz}, \textsf{PDG} \cite{ParticleDataGroup:2008zun}, and \textsf{SLD} \cite{SLD:1998coh,SLD:2003ogn} totaling 4185 data points. The results of the tuning are shown in Tab.~\ref{tab:tune} where we can see that the best-fit value of {\tt probQQtoQ} is 0.111 which yields a deviation of 1-$1.5\sigma$ from the \textsf{HFLAV} estimation of the $b$-baryon fragmantation function but only $1\sigma$ away from the \textsf{DELPHI} measurement of $\langle n_{\Lambda_b^0} \rangle$. The corresponding goodness-of-fit per degrees of freedom at the minimum is $\chi^2/N_{\rm df} = 4943.16/4171 \approx 1.18$. In Fig. \ref{fig:tune:correlation} we show the correlation matrix at the best-fit point where we can see that {\tt probQQtoQ} has a small correlation with other parameters with the maximum being $50\%$ with {\tt mesonUDvector}. 
We must stress out that due to limited accuracy in both the theoretical modeling and the experimental measurements of {\it e.g.} baryon rates, large theoretical uncertainties maybe assessed in our framework. These uncertainties can be estimated using different methods and in particular the Hessian method, widely used in particle-physics community, is the main method of the \textsf{Professor} toolkit. In essence, the uncertainties are estimated by diagonalizing the covariance matrix near the minimum -- best-fit region --. The second-order term of the expansion of $\Delta \chi^2$ distribution, $H_{ij} = \partial^2 \chi^2/\partial x_i \partial x_j$ is called the Hessian matrix whose diagonal form leads to the principal directions or eigenvectors and the corresponding poles are the eigenvalues. Therefore, for a set of $N_{\rm params}$ we get $2 \times N_{\rm params}$ variations; which are 28 variations in one our case. The size of the variations is penalized by a constraint on the corresponding hypersphere of $2 \times N_{\rm params}$ dimensions called the tolerance (see Ref. \cite{Jueid:2023vrb} for a detailed discussion).  A one-sigma uncertainty (eigentunes) is found for $\Delta\chi^2/N_{\rm df} = 1$ while a $2\sigma$ eigentunes corresponds to $\Delta\chi^2 = 4$ where $\Delta\chi^2 \equiv \chi^2 - \chi^2_{\rm min}$.

\section{Coalescence model and calibration of $p_c$ to \textsf{ALICE} data}
\label{appx:ALICEcalib}

\subsection{Monte Carlo setup}

We use \textsf{Pythia 8.311} \cite{Bierlich:2022pfr} to generate Monte Carlo (MC) events for antinuclei generated from DM annihilation or $pp$ collisions. In particular, for DM annihilation, \textsf{Pythia} generate the annihilation of one positron and one electron with an head-on collision with a center of mass energy equal to twice the DM mass.
Instead, for $pp$ collision we fix the center of mass energy to 7 TeV since we use the data for $\bar{p}$, $\overline{\text{D}}$ and ${}^3\overline{\text{He}}$ from Ref.~\cite{PhysRevC.97.024615}.
We simulate $10^{10}$ events for spectra from DM annihilations and $10^9$ from $pp$ collisions to have sufficient statistics.

For each \textsf{Pythia} simulation we select all the pairs of $\bar{n}$ and $\bar{p}$ present in the event list and determine their difference of momenta $\Delta p$ and of distance $\Delta r$ calculated in the reference frame of the $\bar{n}$ and $\bar{p}$ pair.
Then, we apply the coalescence criteria discussed in the manuscript main text, \textit{i.e.}~we test if $\Delta r<3$ fm and $\Delta p<p_c$, where $p_c$ is the coalescence momentum.
If the coalescence criterion is satisfied for a pair of $\bar{n}$ and $\bar{p}$, we assume that the $\overline{\rm D}$ is formed and we calculate its kinetic energy in the center-of-mass reference frame. Once the simulations are finished, we calculate the spectrum from DM annihilations as:
\begin{equation}
\frac{dN_{\rm{DM}}}{dK_i} = \frac{N_i (K\in[K_i,K_i+\Delta K])}{\Delta K},
\end{equation}
where $dN/dK_i$ represents the spectrum evaluated for the $i$-th bin with kinetic energy between $[K_i,K_i+\Delta K]$.
Instead, for $pp$ collisions we select only the events detected at midrapidity, \textit{i.e.}~with $|y|<0.5$ (following Ref.~\cite{PhysRevC.97.024615}), and evaluate the spectra as:
\begin{equation}
\frac{dN_{\rm{pp}}}{p_{T,i}} = \frac{N_i (K\in[p_{T,i},p_{T,i}+\Delta p_T])}{\Delta p_T},
\end{equation}
where $dN/dp_{T,i}$ represents the spectrum evaluated for the $i$-th bin with kinetic energy between $[p_{T,i},p_{T,i}+\Delta p_T]$.

%%%%%%%%%%%%%%%%%%%%%%%%%%%%%%%%%%%%%%%%%%%%%%%%%%%%%%%%%%%%%%%%%%%%%%%%%%
\subsection{Tuning of the coalescence model for ${}^3\overline{\rm He}$ production}
%%%%%%%%%%%%%%%%%%%%%%%%%%%%%%%%%%%%%%%%%%%%%%%%%%%%%%%%%%%%%%%%%%%%%%%%%%

One of the main goal of the paper is to predict the spectrum of ${}^3\overline{\text{He}}$ and the possible contribution of the $b$-baryon weak decays to its yield.
However, we cannot use $e^{\pm}$ data, which are relevant for predictions of particle produced from DM annihilations, because no ${}^3\overline{\text{He}}$ event has ever been measured by any \textsf{LEP} experiment.
Therefore, we decide to calibrate the coalescence model for the rate of ${}^3\overline{\text{He}}$ by using the measurement of \textsf{ALICE} in $pp$ collisions at $\sqrt{s} = 7$ TeV \cite{PhysRevC.97.024615} and check a posteriori if the coalescence model found is compatible with the one found in Ref.~\cite{DiMauro:2024kml} for $\overline{\text{D}}$ production at \textsf{ALEPH}.

\begin{figure*}[!ht]
\includegraphics[width=0.49\linewidth]{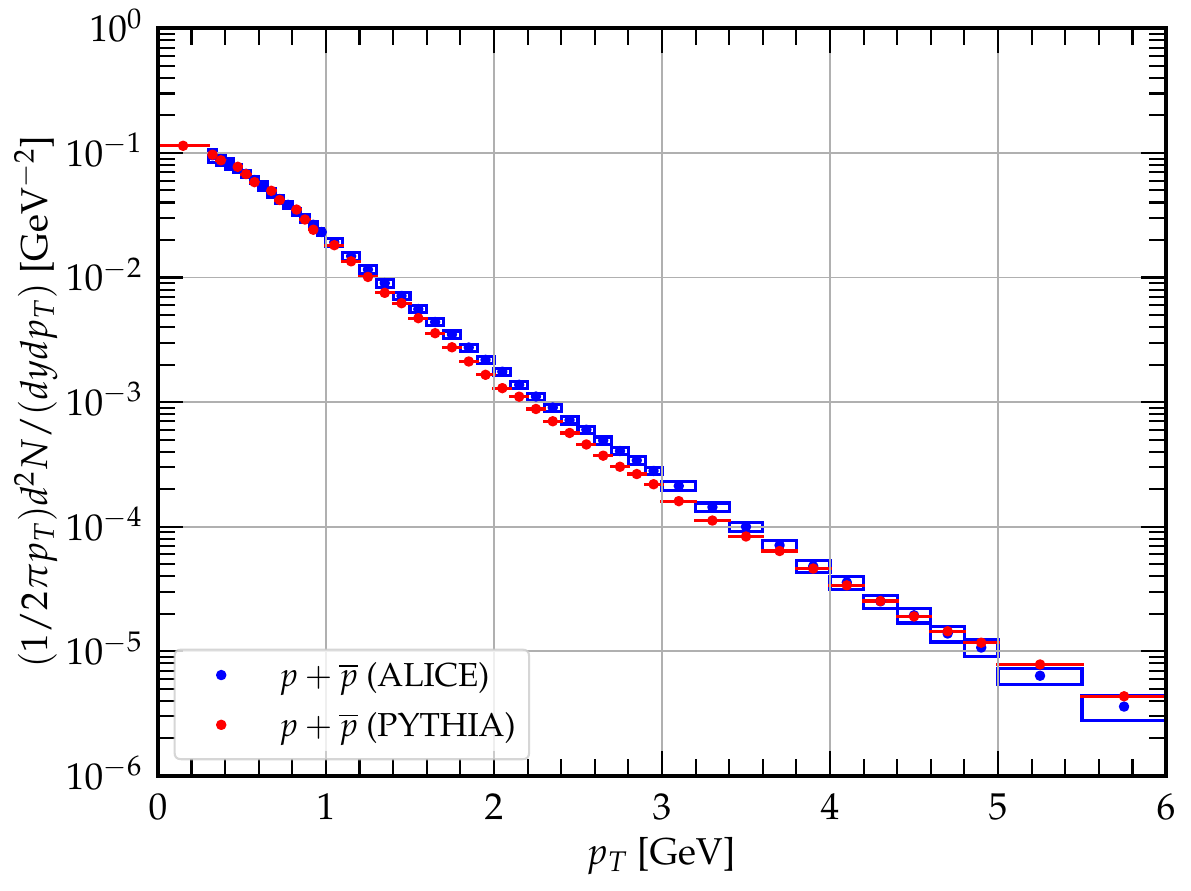}
\includegraphics[width=0.49\linewidth]{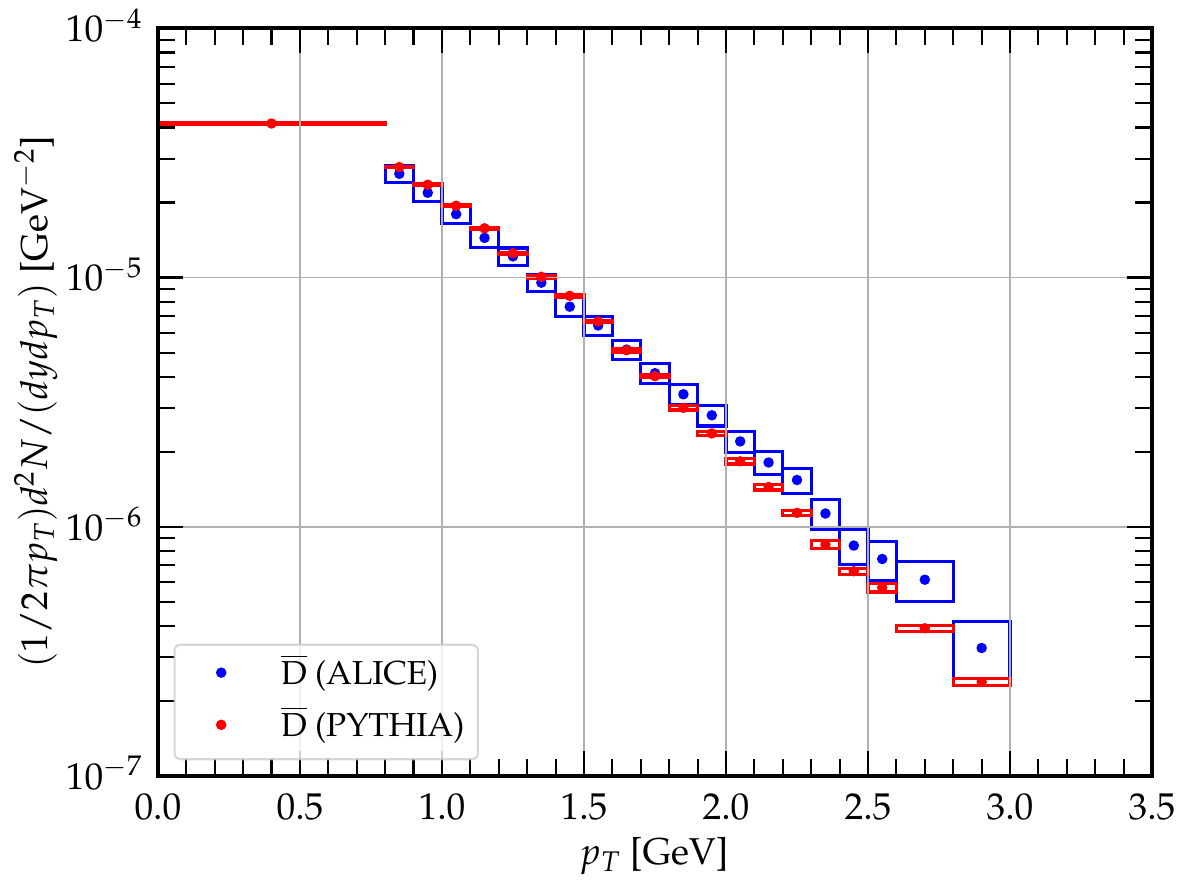}
\includegraphics[width=0.49\linewidth]{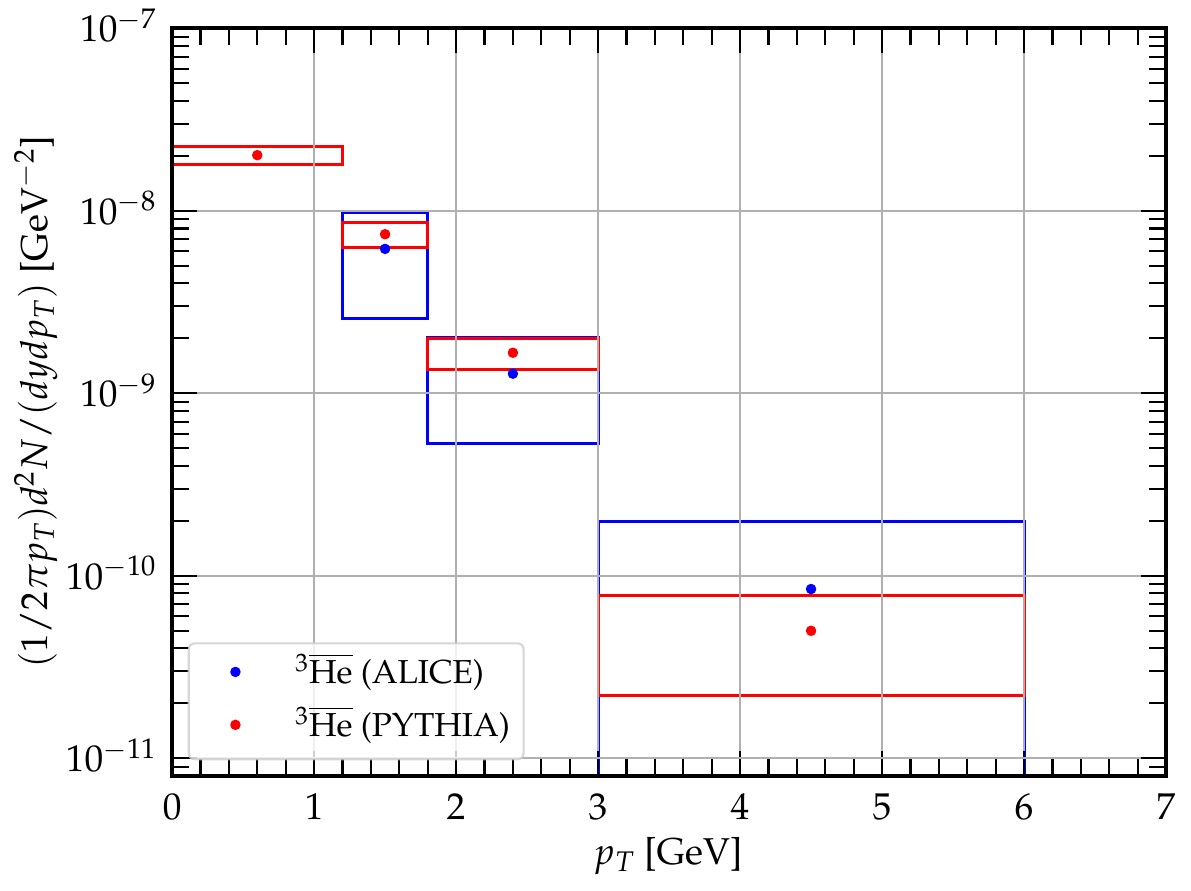}
    \caption{Source spectra for the production of $p+\bar{p}$ (top left), $\overline{\text{D}}$ (top right panel) and ${}^3\overline{\text{He}}$ (bottom panel) from $pp$ collision at $\sqrt{s}=7$ TeV obtained with our tune of \textsf{Pythia} and our coalescence model, for our best-fit value of the coalescence momentum $p_c = 0.20$ GeV. We also show the \textsf{ALICE} data taken at the same center of mass energy from \cite{PhysRevC.97.024615}.}
    \label{fig:Alice}
\end{figure*}

\begin{figure*}[!ht]
\includegraphics[width=0.49\linewidth]{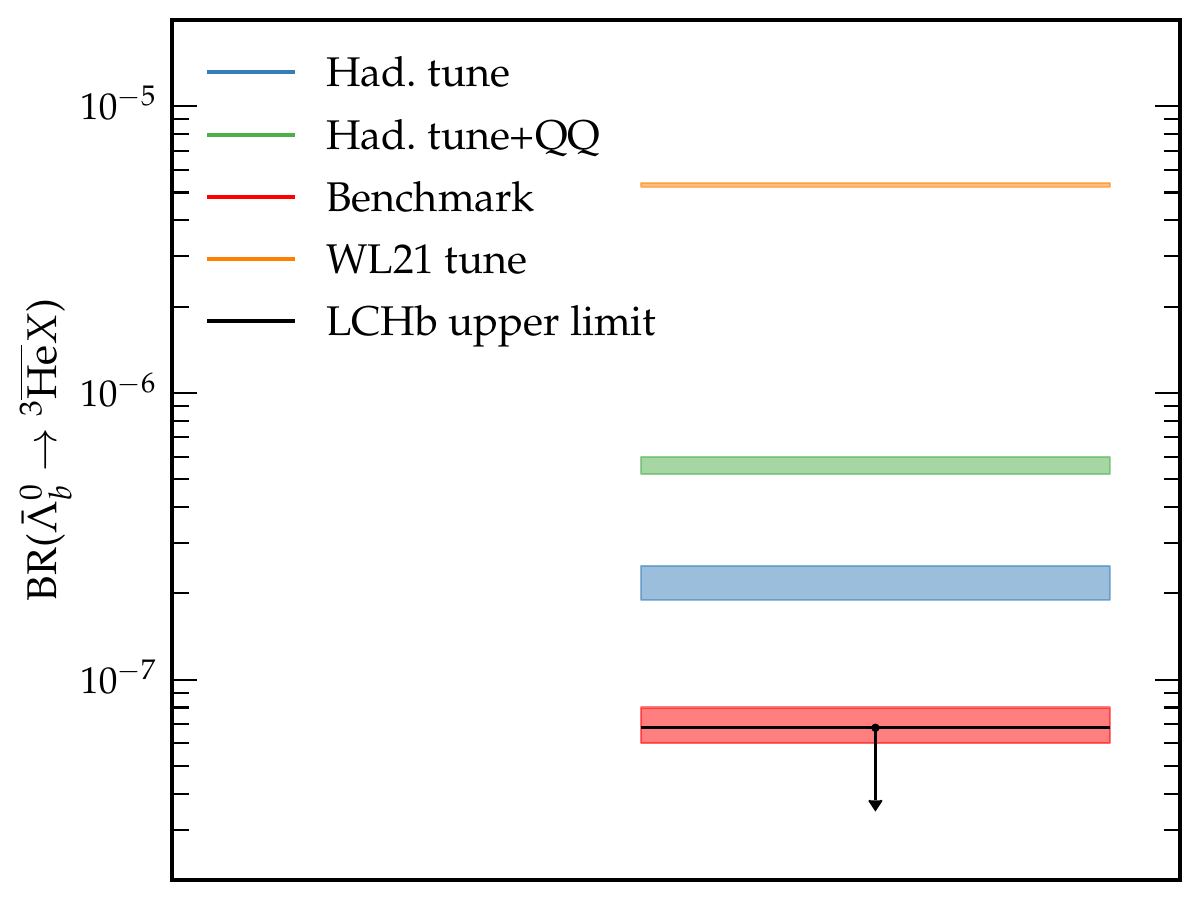}
    \caption{This figure shows the estimate for $\operatorname{BR}\left(\bar{\Lambda}^0_b \rightarrow {}^3\overline{\text{He}}X\right)$ obtained with the different models tested in the paper along with the upper limit measured by LCHb \cite{Moise:2024wqy}.}
    \label{fig:BrUL}
\end{figure*}

The \textsf{Pythia} tunes available for the production of baryons from $pp$ collisions are different from the one generated for $e^{\pm}$. We thus have to use for $pp$ collisions a \textsf{Pythia} tune that is different from the one we found by fitting \textsf{LEP} data. 
We thus derive first a specific hadronization model setup that properly predicts the proton yield measured by \textsf{ALICE}. We find that by using the default \textsf{Monash} tune -- {\tt Tune:pp = 4} -- and the following parameter choices; {\tt StringZ:aLund = 0.360}, {\tt StringZ:bLund = 0.560}, and {\tt StringFlav:ProbStoUD = 0.200}, the predicted proton multiplicity is $1\sigma$ compatible with the \textsf{ALICE} measurement \cite{PhysRevC.97.024615}. In the top left panel of Fig.~\ref{fig:Alice} we show the comparison of our \textsf{ALICE} tune and the data for $p+\bar{p}$.

Once we calibrate the antiprotons spectra, we can refine our coalescence model. In order to do this, we fit the \textsf{ALICE} ${}^3\overline{\text{He}}$ yield finding that for $\Delta r< 3$ fm the best-fit coalescence momentum that is $0.20 \pm 0.01$ GeV. Note that this value is compatible within $1\sigma$ with the value of the coalescence momentum we obtained using the \textsf{ALEPH} data \cite{DiMauro:2024kml} that was $0.21\pm0.02$ GeV. We show in Fig.~\ref{fig:Alice} the comparison of the $\overline{\text{D}}$ and ${}^3\overline{\text{He}}$ spectra with \textsf{ALICE} data. 
It is quite promising that the coalescence model used to fit the ${}^3\overline{\text{He}}$ data also works remarkably well to predict the right yield $\overline{\text{D}}$.
The value of $p_c=0.20$ GeV found with the fit to the ${}^3\overline{\text{He}}$ spectra from \textsf{ALICE} data is used in the Letter.

%%%%%%%%%%%%%%%%%%%%%%%%%%%%%%%%%%%%%%%%%%%%%%%%%%%%%%%%%%%%%%%%%%%%%
\section{Decay channels of $\Lambda_b^0$}
\label{app:Brlambdab}
%%%%%%%%%%%%%%%%%%%%%%%%%%%%%%%%%%%%%%%%%%%%%%%%%%%%%%%%%%%%%%%%%%%%%

%channel: onMode="1" bRatio="0.0200000" products="3122 311 211 211 -211 -211"
%channel: onMode="1" bRatio="0.0120000" meMode="22" products="-2 1 2 2101"
%channel: onMode="1" bRatio="0.4411147" meMode="23" products="-2 1 4 2101"
%channel: onMode="1" bRatio="0.0910000" meMode="43" products="-2 4 1 2101"
%channel: onMode="1" bRatio="0.0120000" meMode="22" products="-4 3 2 2101"
%channel: onMode="1" bRatio="0.0800000" meMode="43" products="-4 3 4 2101"

%particle: id="1" name="d" antiName="dbar" spinType="2" chargeType="-1" colType="1" m0="0.33000"
%particle: id="2" name="u" antiName="ubar" spinType="2" chargeType="2" colType="1" m0="0.33000"
%particle: id="3" name="s" antiName="sbar" spinType="2" chargeType="-1" colType="1" m0="0.50000"
%particle: id="4" name="c" antiName="cbar" spinType="2" chargeType="2" colType="1" m0="1.50000"
%particle: id="5" name="b" antiName="bbar" spinType="2" chargeType="-1" colType="1" m0="4.80000"
%particle: id="6" name="t" antiName="tbar" spinType="2" chargeType="2" colType="1" m0="173.00000" mWidth="1.40000" mMin="86.00000" mMax="0.00000"

The $\Lambda^0_b$ is a $b$-baryon with a rest mass of 5620 MeV, a decay time of about 1.4 ps and it is made of $u \, d \, b$ valence quarks.
In this section we will talk about the decay of the particles but we remind the reader that we are interested in its antiparticle ($\bar{\Lambda}^0_b$).
A $\Lambda^0_b$ of 20 GeV of total energy would travel on average about 2 cm before decaying as expected by weak decays. This happens also for the other $b$-baryons whose decays are typically labeled as display-vertexes decays. 
The most frequent and well measured decay channel of the $\Lambda^0_b$ is the one into $\Lambda_c^+ \ell^- \overline{\nu}_\ell$ which happens about $6\%$ of the times. We report in Tab.~\ref{tab:lambdabdecays} the list of the most relevant decay channels.
However, this decay mode is not of interest for the production of $\overline{\text{D}}$ and ${}^3\overline{\text{He}}$.

As outlined in the main text, \textsf{Pythia} takes into account the rare decays of the $\Lambda^0_b$ into $\bar{p}\bar{n}$ pairs by considering three quarks plus one di-quark states.
The most important decay channel is:
%-2 1 4 2101
\begin{equation}
\Lambda^0_b \rightarrow \bar{u} \, d \, c \,  (ud)_0,
\end{equation}
where $(ud)_0$ is the di-quark state. This decay channel happens about $53\%$ of the times as implemented in \textsf{Pythia}.
Instead the decay:
\begin{equation}
\Lambda^0_b \rightarrow \bar{u} \, u \, d \,  (ud)_0,
\end{equation}
which is more relevant for the antinuclei production, is $1.2\%$ frequent. 
As explained in the main text, the difference of the two channel branching ratios is mostly related to the values $\left|V_{u b}\right|$ and $\left|V_{c b}\right|$ of the Cabibbo–Kobayashi–Maskawa matrix, which controls the transition probability for $b\rightarrow u$ and $b\rightarrow c$, where. In particular, $\left|V_{u b}\right|^2 /\left|V_{c b}\right|^2 \approx 100$.
The remaining difference is explained by the fact that the allowed phase space prefers a higher multiplicity in $b \rightarrow u$ than in $b \rightarrow c$. 
We report in Tab.~\ref{tab:Brtuning} the complete list of the di-quark modes with the default \textsf{Pythia} values.
Such branching ratios are however not measured by any experiment yet but can be {\it thought} as external input parameters in \textsf{Pythia} whose values are just guessed and thus subject to large theoretical uncertainties.

We show in Tab.~\ref{tab:lambdabdecays} the most relevant $\Lambda_b^0$ decay channels with the measurements reported in the \textsf{PDG} \cite{ParticleDataGroup:2024cfk} and the predictions from the default \textsf{Pythia} and our modified branching ratios.
We can see that \textsf{Pythia} matches well the measurements except for the channel  $\operatorname{BR}\left(\bar{{\Lambda}}_{\mathrm{b}}^0 \rightarrow \Lambda_{\mathrm{c}}^{-} p \bar{p} \pi^{+}\right)$ for which it predicts a too large value. As for the $\Lambda_c^+$ the default \textsf{Pythia} does not predict the right inclusive $\bar{p}/\bar{n}$ decays while our tune model for the branching ratios into quark-diquark states matches the data remarkably well. 

The modifications we have applied to the $\Lambda_b^0$ decay modes have been dictated by the \textsf{LHCb} upper limit for the branching ratio of the process $\operatorname{Br}\left(\bar{\Lambda}_b^0 \rightarrow {}^3\overline{\text{He}}X\right)$, which is $6.3\times 10^{-8}$.
In Tab.~\ref{tab:Brtuning} we report the default and the modified \textsf{Pythia} values of the $\Lambda_b^0$ decay channels into di-quark states. We can see that we only slightly modified the ones for the $\Lambda_b^0$. The main change we applied is a reduction of the $\bar{u} \, d \, u (ud)_0$ from the default 0.12 to 0.001. This is indeed the decay mode that contributes the most to the antinucleons production. With the tuned values of the decay channels we find $\operatorname{Br}\left(\bar{\Lambda}_b^0 \rightarrow {}^3\overline{\text{He}}X\right) = (7 \pm 1) \times 10^{-8}$, which is compatible at $1\sigma$ with the \textsf{LHCb} upper limit.
We show in Fig.~\ref{fig:BrUL} the comparison of the theoretical values for $\operatorname{Br}\left(\bar{\Lambda}_b^0 \rightarrow {}^3\overline{\text{He}}X\right)$ obtained with our models compared with the \textsf{LHCb} upper limit. In particular, we see that the effect of increasing the {\tt probQQtoQ} parameter from 0.11 to 0.19, i.e.~between the models {\bf Had.tune} and {\bf Had.tune+QQ}, is an higher branching ratio by a factor of almost 3.

%For the $\Lambda_c^+$, the \textsf{Pythia} default does not provide the correct branching ratios into $pX$ and $nX$. To address this, we also modify the branching ratios of $\Lambda_c^+$ into di-quark modes. Using the values reported in Tab.~\ref{tab:Brtuning}, we find a good match for $\operatorname{BR}\left(\Lambda_c^+ \rightarrow pX, nX\right)$ to the measurements, as shown in Tab.~\ref{tab:lambdabdecays}, which also reports these branching ratios for different \textsf{Pythia} discussed in the following section. \\

\begin{table}[H]
\setlength\tabcolsep{8pt}
%  \begin{center}
    \centering
    \begin{tabular}{l c c}
        \toprule \toprule
        $\Lambda_b^0$ decay mode & Default \textsf{Pythia} & Tuned \textsf{Pythia} \\
        \toprule
%        \hline
        $\overline{u}du(ud)_0$ & $0.012$ & $0.001$ \\
%        \hline
        $\overline{u}dc(ud)_0$ & $0.5321147$ & $0.5321147$ \\
%        \hline
        $\overline{c}su(ud)_0$ & $0.012$ & $0$ \\
%        \hline
        $\overline{c}sc(ud)_0$ & $0.08$ & $0.103$ \\
       %\toprule
       %$\Lambda_c^+$ decay mode & & \\
       %\toprule
       %$d (uu)_1$ & $0.03$ & $0.346$ \\
       %$s (uu)_1$ & $0.095$ & $0.346$ \\
       \bottomrule \bottomrule
    \end{tabular}
    \caption{Tune of the branching ratios of $\Lambda_b^0$ into di-quark states. For the latter, the remaining branching ratios have been set to $0$.}
    \label{tab:Brtuning}
\end{table}

\begin{table}[H]
\setlength\tabcolsep{9pt}
%  \begin{center}
    \centering
    \begin{tabular}{l c c c c}
        \toprule \toprule
        $\bar{\Lambda}_b^0$ decay mode & Measured BR & WL21 tune & Had.~tune & Benchmark \\
        \toprule
        $\Lambda_c^-\ell^+ \nu_\ell$ & $(6.2^{+1.4}_{-1.3})\%$ & $(5.9569\pm0.0009)\%$ & $(5.956\pm0.002)\%$ & $(5.957\pm0.001)\%$ \\
        $\bar{p}\overline{D}^0\pi^+$ & $(6.2\pm0.6)\times10^{-4}$ & $(7.13\pm0.01)\times10^{-4}$ & $(6.11\pm0.02)\times10^{-4}$ & $(6.64\pm0.01)\times10^{-4}$ \\
        $\bar{p}D^-\pi^-\pi^+$ & $(2.7\pm0.4)\times10^{-4}$ & $(2.544\pm0.006)\times10^{-4}$ & $(2.154\pm0.008)\times10^{-4}$ & $(2.215\pm0.006)\times10^{-4}$ \\
        $\bar{p}\pi^+$ & $(4.6\pm0.8)\times10^{-6}$ & $(1.219\pm0.005)\times10^{-4}$ & $(1.035\pm0.006)\times10^{-5}$ & $(1.41\pm0.02)\times10^{-5}$ \\
        $\Lambda_c^-K^+K^-\pi^+$ & $(1.02\pm0.11)\times10^{-3}$ & $(1.342\pm0.002)\times10^{-3}$ & $(1.391\pm0.003)\times10^{-3}$ & $(1.354\pm0.002)\times10^{-3}$ \\
        $\Lambda_c^-p\overline{p}\pi^+$ & $(2.63\pm0.23)\times10^{-4}$ & $(3.285\pm0.003)\times10^{-3}$ & $(2.677\pm0.009)\times10^{-3}$ & $(4.615\pm0.009)\times10^{-4}$ \\
        $\bar{p}\bar{n}X$ & -- & $(5.337\pm0.003)\times10^{-3}$ & $(7.32\pm0.02)\times10^{-4}$ & $(6.56\pm0.01)\times10^{-4}$ \\
        $\bar{p}\bar{p}\bar{n}X$ & -- & $(7.64\pm0.04)\times10^{-5}$ & $(1.04\pm0.06)\times10^{-6}$ & $(2.4\pm0.3)\times10^{-7}$ \\
        ${}^3\overline{\text{He}}X$ & $< 6.8\times10^{-8}$ & $(5.32\pm0.08)\times10^{-6}$ & $(2.2\pm0.3)\times10^{-7}$ & $(7\pm1)\times10^{-8}$ \\
        %\toprule
        %$\Lambda_c^-$ decay mode & & & & \\
        %\toprule
        %$\bar{p}X$ & $(50\pm16)\%$ & $(14.254\pm0.002)\%$ & $(14.404\pm0.003)\%$ & $(41.250\pm0.003)\%$ \\
        %$\bar{n}X$ & $(32.6\pm1.6)\%$ & $(4.9004\pm0.0008)\%$ & $(4.909\pm0.001)\%$ & $(34.206\pm0.003)\%$ \\
        \bottomrule \bottomrule
    \end{tabular}
    \cprotect\caption{Branching ratios of $\bar{\Lambda}_b^0$ into diverse modes. The first column reports their measurement, provided by \textsf{PDG} \cite{ParticleDataGroup:2024cfk} and \textsf{LHCb} \cite{Moise:2024wqy}. The results were obtained using $10^{10}$ events in \textsf{Pythia}.}
    \label{tab:lambdabdecays}
\end{table}

\end{document}